\RequirePackage{fix-cm} 
\documentclass[a4paper, twoside, reqno, 12pt]{amsart}
\usepackage{fixltx2e}   

\usepackage{etex}

\usepackage[latin1]{inputenc}
\usepackage[T1]{fontenc}

\usepackage{esint}
\usepackage{dsfont}
\usepackage{xspace}
\usepackage{amsgen}
\usepackage{amsthm}
\usepackage{amssymb}
\usepackage{amsmath}
\usepackage{wasysym}
\usepackage{upgreek}
\usepackage{amsfonts}
\usepackage{stmaryrd}
\usepackage{mathtools}
\usepackage[mathcal, mathscr]{euscript}

\usepackage{mathrsfs}
\DeclareMathAlphabet{\mathscrbf}{OMS}{mdugm}{b}{n}

\usepackage{multicol}
\usepackage{multirow}

\usepackage{a4wide}

\usepackage[dvipsnames, table]{xcolor}
\definecolor{bckg}{RGB}{20.8, 20.8, 20.8}
\definecolor{oneblue}{rgb}{0.0, 0.0, 0.85}
\definecolor{Lightblue}{RGB}{214, 214, 214}
\definecolor{bluepigment}{rgb}{0.2, 0.2, 0.6}
\definecolor{charcoal}{rgb}{0.21, 0.27, 0.31}
\definecolor{denimblue}{rgb}{0.08, 0.38, 0.74}
\definecolor{Lightgray}{rgb}{0.89, 0.89, 0.89}
\definecolor{darkgrey}{rgb}{0.273, 0.281, 0.30}
\definecolor{darkelectricblue}{rgb}{0.33, 0.41, 0.47}

\usepackage[sort&compress, comma, square, numbers]{natbib}

\usepackage{psfrag}
\usepackage{subfig}
\usepackage{graphicx}
\usepackage{subfloat}
\usepackage{morefloats}
\usepackage{indentfirst}

\usepackage{acronym}
\usepackage{microtype}
\usepackage[labelsep=period,%
            labelfont={bf,sf,color=bluepigment},%
            justification=raggedright]{caption}

\usepackage[perpage, symbol]{footmisc}

\usepackage[usenames, pdf]{pstricks}
\usepackage{epsfig}
\usepackage{pst-grad} 
\usepackage{pst-plot} 

\usepackage{url}
\usepackage[colorlinks,
           urlcolor=oneblue,
           linkcolor=denimblue,
           citecolor=NavyBlue,
           bookmarksopen=false,
           pdfpagemode=UseNone,
           pagebackref]{hyperref}

\usepackage[explicit]{titlesec}

\titleformat{\section}[block]
  {\color{NavyBlue}\Large\sffamily\bfseries}
  {}
  {0.0em}
  {\colorbox{bckg!5}{\strut\parbox{\dimexpr\linewidth-2\fboxsep\relax}{\thesection. #1}}}
  [\vspace*{0.33em}]

\titleformat{name=\section,numberless}[block]
  {\color{NavyBlue}\Large\sffamily\bfseries}
  {}
  {0.0em}
  {\colorbox{bckg!5}{\strut\parbox{\dimexpr\linewidth-2\fboxsep\relax}{#1}}}
  [\vspace*{0.33em}]

\titleformat{\subsection}
  {\color{NavyBlue}\large\sffamily\bfseries}
  {}
  {0.0em}
  {\colorbox{bckg!5}{\parbox{\dimexpr\linewidth-2\fboxsep\relax}{\thesubsection. #1}}}
  [\vspace*{0.33em}]

\titleformat{name=\subsection,numberless}
  {\color{NavyBlue}\Large\sffamily\bfseries}
  {}
  {0em}
  {\colorbox{bckg!5}{\parbox{\dimexpr\linewidth-2\fboxsep\relax}{#1}}}
  [\vspace*{0.33em}]

\titleformat{\subsubsection}
  {\color{bluepigment}\sffamily\normalsize\bfseries}
  {\thesubsubsection}
  {0.5em}
  {#1}
  [\vspace*{0.33em}]

\titleformat{\paragraph}[runin]
  {\color{bluepigment}\sffamily\small\bfseries}
  {}
  {0em}
  {#1}

\titlespacing{\section}{0.0em}{1.5em plus 2pt minus 2pt}%
{1.0em plus 2pt minus 2pt}[0em]
\titlespacing{\subsection}{0.5em}{1.5em plus 2pt minus 2pt}%
{1.0em}[0em]
\titlespacing{\subsubsection}{0.5em}{1.5em plus 2pt minus 2pt}%
{1.0em plus 2pt minus 2pt}[0em]

\usepackage{titletoc}

\contentsmargin{0.5em}
\newlength{\tocsep} 
\titlecontents{section}[\tocsep]
  {\addvspace{10pt}\bfseries\sffamily}
  {\contentslabel[\thecontentslabel]{\tocsep}}
  {}
  {\ \titlerule*[0.75pc]{.}\ \thecontentspage}
  []
\titlecontents{subsection}[\tocsep]
  {\addvspace{8pt}\sffamily}
  {\contentslabel[\thecontentslabel]{\tocsep}}
  {}
  {\ \titlerule*[0.5pc]{.}\ \thecontentspage}
  []
\titlecontents*{subsubsection}[\tocsep]
  {\addvspace{2pt}\footnotesize\sffamily}
  {}
  {}
  {\ \titlerule*[0.35pc]{.}\ \thecontentspage}
  [\\*]

\makeatletter
\def\@setauthors{%
  \begingroup
  \def\thanks{\protect\thanks@warning}%
  \trivlist
  \centering\footnotesize \@topsep30\p@\relax
  \advance\@topsep by -\baselineskip
  \item\relax
  \author@andify\authors
  \def\\{\protect\linebreak}%
  \textsc{\normalsize\textcolor{darkelectricblue}{\authors}}%
  \ifx\@empty\contribs
  \else
    ,\penalty-3 \space \@setcontribs
    \@closetoccontribs
  \fi
  \endtrivlist
  \endgroup
}
\def\@settitle{\begin{center}%
  \baselineskip14\p@\relax
    \bfseries
    \textsc{\Large\textcolor{charcoal}{\@title}}
  \end{center}%
}
\makeatother

\usepackage{enumitem}
\setlist[description]{%
  topsep=30pt,               
  itemsep=5pt,               
  font={\bfseries\sffamily\color{NavyBlue}}, 
}

\usepackage{fancyhdr}
\usepackage{lastpage}

\newcommand*\Title{\textcolor{bluepigment}{On the optimal experimental design}}
\newcommand*\Authors{\textcolor{bluepigment}{J.~Berger, D.~Dutykh \& N.~Mendes}}
\newcommand*{\plogo}{\textcolor{gray}{{\texttt{arXiv.org} / \textsc{hal}}}} 

\numberwithin{equation}{section}

\newcommand{\IP}{\text{IP}}
\newcommand{\vP}{\mathbf{P}}
\newcommand{\vU}{\mathbf{U}}
\newcommand{\vX}{\mathbf{X}}
\newcommand{\e}[1]{\times 10^{#1}}
\newcommand\eqs[1]{Eqs. \eqref{#1}}

\newcommand{\eg}{\emph{e.g.}\/ }

\usepackage{acronym}
\acrodef{bvp}[BVP]{Boundary Value Problem}
\acrodef{NSWE}{Nonlinear Shallow Water Equations}

\begin{document}

\title[\Title]{On the optimal experimental design for heat and moisture parameter estimation}

\author[J.~Berger]{Julien Berger$^*$}
\address{Thermal Systems Laboratory, Mechanical Engineering Graduate Program, Pontifical Catholic University of Paran\'a, Rua Imaculada Concei\c{c}\~{a}o, 1155, CEP: 80215-901, Curitiba -- Paran\'a, Brazil}
\email{Julien.Berger@pucpr.edu.br}
\urladdr{https://www.researchgate.net/profile/Julien\_Berger3/}
\thanks{$^*$ Corresponding author}

\author[D.~Dutykh]{Denys Dutykh}
\address{LAMA, UMR 5127 CNRS, Universit\'e Savoie Mont Blanc, Campus Scientifique, 73376 Le Bourget-du-Lac Cedex, France}
\email{Denys.Dutykh@univ-savoie.fr}
\urladdr{http://www.denys-dutykh.com/}

\author[N.~Mendes]{Nathan Mendes}
\address{Thermal Systems Laboratory, Mechanical Engineering Graduate Program, Pontifical Catholic University of Paran\'a, Rua Imaculada Concei\c{c}\~{a}o, 1155, CEP: 80215-901, Curitiba -- Paran\'a, Brazil}
\email{Nathan.Mendes@pucpr.edu.br}
\urladdr{https://www.researchgate.net/profile/Nathan\_Mendes/}



\begin{titlepage}
\thispagestyle{empty} 
\noindent
{\Large Julien \textsc{Berger}}\\
{\it\textcolor{gray}{Pontifical Catholic University of Paran\'a, Brazil}}
\\[0.02\textheight]
{\Large Denys \textsc{Dutykh}}\\
{\it\textcolor{gray}{CNRS--LAMA, Universit\'e Savoie Mont Blanc, France}}
\\[0.02\textheight]
{\Large Nathan \textsc{Mendes}}\\
{\it\textcolor{gray}{Pontifical Catholic University of Paran\'a, Brazil}}
\\[0.08\textheight]

\vspace*{1.1cm}

\colorbox{Lightblue}{
  \parbox[t]{1.0\textwidth}{
    \centering\huge\sc
    \vspace*{0.7cm}
    
    \textcolor{bluepigment}{On the optimal experimental design for heat and moisture parameter estimation}
    
    \vspace*{0.7cm}
  }
}

\vfill 

\raggedleft     
{\large \plogo} 
\end{titlepage}


\newpage
\thispagestyle{empty} 
\par\vspace*{\fill}   
\begin{flushright} 
{\textcolor{denimblue}{\textsc{Last modified:}} \today}
\end{flushright}


\newpage
\maketitle
\thispagestyle{empty}


\begin{abstract}

In the context of estimating material properties of porous walls based on in-site measurements and identification method, this paper presents the concept of Optimal Experiment Design (OED). It aims at searching the best experimental conditions in terms of quantity and position of sensors and boundary conditions imposed to the material. These optimal conditions ensure to provide the maximum accuracy of the identification method and thus the estimated parameters. The search of the OED is done by using the \textsc{Fisher} information matrix and a priori knowledge of the parameters. The methodology is applied for two case studies. The first one deals with purely conductive heat transfer. The concept of optimal experiment design is detailed and verified with 100 inverse problems for different experiment designs. The second case study combines a strong coupling between heat and moisture transfer through a porous building material. The methodology presented is based on a scientific formalism for efficient planning of experimental work that can be extended to the optimal design of experiments related to other problems in thermal and fluid sciences.

\bigskip
\noindent \textbf{\keywordsname:} inverse problem; parameter estimation; Optimal Experiment Design (OED); heat and moisture transfer; sensitivity functions \\


\end{abstract}


\newpage
\tableofcontents
\thispagestyle{empty}


\newpage
\section{Introduction}

Heating or cooling strategies for buildings are commonly based on numerical building physics mathematical models, which can be calibrated using on-site measurements for estimating the properties of the materials constituting the walls, reducing the discrepancies between model predictions and real performance.

Several experimental works at the scale of the wall can be reported from the literature. Instrumented test cells, as ones presented in \cite{Rafidiarison2015, Talukdar2007, James2010, Kalamees2003, Lelievre2014, Desta2011, Stephan2014, Colinart2016} provide measured dataset as temperature and relative humidity at different points in the wall for given boundary conditions. Some experiments at the scale of the whole-building are described in \cite{Labat2015, Cantin2010}. These data can be used to estimate the properties (transport and capacity coefficients) of the materials as reported for instance in \cite{Tasca-Guernouti2011, Nassiopoulos2013} for heat transfer and in \cite{Rouchier2016, Zaknoune2012} for coupled heat and moisture transfer.

\begin{figure}
\captionsetup{type=figure}
\begin{center}
\def\svgwidth{1.05\textwidth}
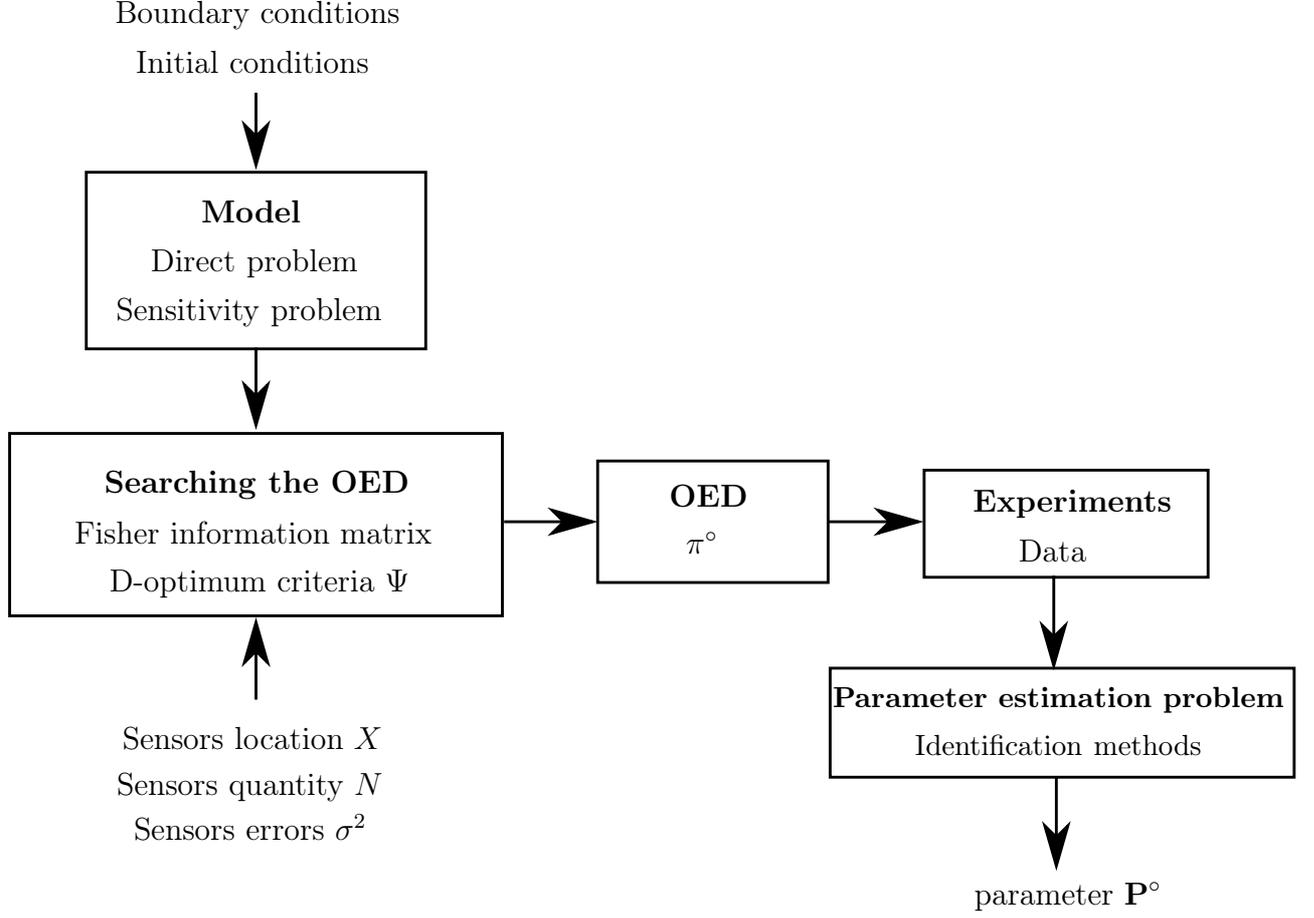
\caption{\small\em Process of searching the Optimal Experiment Design.}
\label{intro_fig:OED}
\end{center}
\end{figure}

The estimation of the unknown parameters $\vP$, \eg wall thermophysical properties, based on observed data and identification methods is illustrated in Figure~\ref{intro_fig:OED}. Observed data are experimentally obtained. The latter is established by a design defining the boundary and initial conditions of the material, the type of sensors as well as their quantity and location. Thus, the accuracy of the estimated parameters $\vP$ strongly depends on the experiment design. The choice of the measurement devices, of the boundary and initial conditions have consequences on the estimation of the parameter. Furthermore, due to the correlation between the parameters, multiple local solutions of the estimation problem exist. Hence, one can address the following questions: what are the best experimental conditions to provide the best conditioning of the identification method? In particular, how many sensors are necessary? Where are their best locations? What boundary and initial conditions should be imposed? Can we really choose them?

These issues deal with searching the Optimal Experiment Design (OED) that enables to identify the parameters $\vP$ of a model with a maximum precision. A first objective of the OED is to adjust the conditions of the experiment design in order to maximize the sensitivity of the output field $u$ to parameters $\vP$. A second objective is to find the optimum location and quantity of sensors. The achievement of these objectives enables to determine conditions of the experiments under which the identification of the parameters will have the maximum accuracy.

The search of the OED is based on quantifying the amount of information contained by the observed field $u_{\mathrm{exp}}\,$. For this, \textsc{Fisher} information matrix is used \cite{Karalashvili2015, Ucinski2004, VandeWouwer2000, Sun2005, Fadale1995, Emery1998, Anderson2005, Alifanov1995}, considering both the model sensitivity, the measurement devices and the parameter correlation. The sensitivity of the output field $u$ with respect to the model parameters $\vP$ is calculated, corresponding to the sensitivity of the cost function of the parameter estimation problem. The higher the sensitivity is, the more information is available in the measurement data and the more accurate is the identification of the parameters. Generally speaking, the methodology of searching the ODE is important before starting any experiment aiming at solving parameter estimation problems. It allows choosing with a deterministic approach the conditions of the experiments. Furthermore, it provides information on the accuracy of parameter estimation.

Several works can be reported on the investigation of OED \cite{Artyukhin1985, Karalashvili2015, Ucinski2004, VandeWouwer2000, Sun2005, Fadale1995, Emery1998, Anderson2005, Alifanov1995}. Among them, in \cite{Artyukhin1985}, the OED is planned as a function of the quantity of sensors and their location, for the estimation of boundary heat flux of non-linear heat transfer. In \cite{Karalashvili2015}, the OED is searched for the estimation of transport coefficient in convection-diffusion equation. In \cite{Nenarokomov2005}, the OED is analysed as a function of the boundary heat flux for the identification of the radiative properties of the material. In \cite{Terejanu2012}, the OED is investigated for chemical reactions using a Bayesian approach. However, the application of OED theory for non-linear heat and moisture transfer with application for the estimation of the properties of a wall is relatively rare.

This article presents the methodology of searching the OED for experiments aiming at solving parameter estimation problems. In the first Section, the concept of OED is detailed and verified for an inverse problem of non-linear heat transfer. The computation of the model sensitivity to the parameter is specified. The OED is sought as a function of the quantity and location of sensors as well as the amplitude and the frequency of the heat flux at the material boundaries. Then the OED for the estimation of hygrothermal properties considering non-linear heat and moisture transfer is investigated. Finally, some main conclusions and perspectives are outlined in the last Section. 

\section{Optimal Experiment Design for non-linear heat transfer problem}
\label{sec1:OED_NL_heat}

First, the methodology of searching the OED is detailed for a problem of non-linear heat transfer. A brief numerical example is given to illustrate the results.

\subsection{Physical problem and mathematical formulation}

The physical problem considers an experiment of a one-dimensional transient conduction heat transfer in domains $ x\in \Omega = \left[0, 1\right]$ and $ t \in \left[0, \tau\right]$. The initial temperature in the body is supposed uniform. The surface of the boundary $\Gamma_D = \left\lbrace x=1 \right\rbrace$ is maintained at the temperature $u_{D}$. A time-harmonic heat flux $q$, of amplitude $A$ and frequency $\omega$, is imposed at the surface of the body denoted by $\Gamma_q\left\lbrace x=0 \right\rbrace$. Therefore, the mathematical formulation of the heat conduction problem can be written as:
\begin{subequations}
\label{sec1_eq:direct_problem}
\begin{align}
&  c^\star \frac{\partial u}{\partial t^\star} 
- \frac{\partial }{\partial x^\star} \left( k^\star(u) \frac{\partial}{\partial x^\star} u \right) =0  
&& x^\star \in \Omega, t^\star \in \left]0, \tau \right], \\
& u=u_{D}  
&& x^\star \in \Gamma_D , t>0, \\
& k^\star(u) \frac{\partial u}{\partial x^\star}  =A^\star \sin(2\pi \omega^\star t^\star) 
&& x^\star \in \Gamma_q , t>0, \\
& u=u_0(x^\star) 
&& x^\star \in \Omega,t^\star=0, \\
& k^\star(u)=\left(k^\star_0+k^\star_1u \right)
\end{align}
\end{subequations}
where the following dimensionless quantities are introduced:
\begin{align*}
& x^\star=\frac{x}{L}, & & u=\frac{T}{T_{ref}}, & & u_D = \frac{T_D}{T_{ref}}, 
& & u_0 = \frac{T_0}{T_{ref}}, & & k^\star_0 = \frac{k_0}{k_r}, \\[5mm]
& k^\star_1=\frac{k_1T_{ref}}{k_{ref}}, & & c^\star = \frac{c}{c_{ref}}, 
& &  t_{ref} = \frac{c_{ref} L^2}{k_{ref}}, & & A^\star= \frac{A L}{k_{ref} T_{ref}}, 
& & \omega^\star = \omega t_{ref}
\end{align*}
where $T$ is the temperature, $c$ the volumetric heat capacity, $k_0$ the thermal conductivity and $k_1$ its dependency on temperature, $L$ the linear dimension of the material, $A$ the intensity and $\omega$ the frequency of the flux imposed on surface $\Gamma_q$. Subscripts $ref$ accounts for a characteristic reference value, $D$ for the Dirichlet boundary conditions, zero ($0$) for the initial condition of the problem and superscript $\star$ for dimensionless parameters.

The problem given by \eqs{sec1_eq:direct_problem}(a-e) is a direct problem when all the thermophysical properties, initial and boundary conditions, as well as the body geometry are known. Here, we focus on the inverse problem consisting in estimating one or more parameters of the material properties (as $c$, $k_0$ and/or $k_1$) using the mathematical model \eqs{sec1_eq:direct_problem}(a-e) and a measured temperature data $u_{exp}$ obtained by $N$ sensors placed in the material at $\vX = \left[x_n\right], n \in \left\lbrace 1, \ldots, N \right\rbrace$. The $M$ unknown parameters are here denoted as vector $\vP=\left[ p_m \right],  m \in \left\lbrace 1, \ldots, M \right\rbrace$. Here, the inverse problems are all non-linear, over-determined solved by considering a least-square single-objective function:
\begin{align}
\label{sec1_eq:fitness_function}
J\left[\vP\right]=|| u_{exp}-\mathfrak{T}\left(u\left(x,t,\vP\right)\right)||^2
\end{align} 
where $u$ is the solution of the transient heat conduction problem \eqs{sec1_eq:direct_problem}(a-e). $u_{exp}$ are the data obtained by experiments. They are obtained by $N$ sensors providing a time discrete measure of the field at specified points within the material. $\mathfrak{T}$ is the operator allowing to compare the solution $u$ at the same space-time points where observations $u_{exp}$ are taken. The cost function can be defined in its matrix form, using the ordinary least squares norm, as:
\begin{align*}
J\left[\vP\right] & = \bigl[ \vU_{exp} - \vU\left(\vP\right) \bigr]^{\top} \bigl[ \vU_{exp} - \vU\left(\vP\right) \bigr]
\end{align*}
where $\vU$ is the vector containing the discrete point of the field $u$ for the discrete set of space-time points obtained by the experiments.

For this study, several experiments are distinguished. Ones aim at estimating one parameter: $m=1$ and $\vP=c$, $\vP=k_0$ or $\vP=k_1$. Others aim at estimating a group of $m=2$ or $m=3$ parameters. In this work, the optimal experiment design will be investigated for both cases.

\subsection{Optimal experiment design}
\label{sec1:OED}

Efficient computational algorithms for recovering parameters $\vP$ have already been proposed. Readers may refer to \cite{Ozisik2000} for a primary overview of different methods. They are based on the minimisation of the cost function $J\left[\vP\right]$. For this, it is required to equate to zero the derivatives of $J\left[\vP\right]$ with respect to each of the unknown parameters $p_m$. Associated to this necessary condition for the minimisation of $J\left[\vP\right]$, the scaled dimensionless local sensitivity function \cite{Finsterle2015} is introduced:
\begin{align}\label{sec1_eq:sensitivity_matrix}
  \Theta_m(x,t)= \frac{\sigma_{p}}{\sigma_{u}} \, \frac{\partial u}{\partial p_m} ,\qquad \forall m \in \left\lbrace 1, \ldots, M \right\rbrace
\end{align}
where $\sigma_u$ is the variance of the error measuring $u_{exp}$. The parameter scaling factor $\sigma_p$ equals 1 as we consider that prior information on parameter $p_m$ has low accuracy. It is important to note that all the algorithm have been developed considering the dimensionless problem in order to compare only the order of variation of parameters and observation (and thus avoid units and scales effects).

The sensitivity function $\Theta_m$ measures the sensitivity of the estimated field $u$ with respect to change in the parameter $p_m$ \cite{Nenarokomov2005, Ozisik2000, Artyukhin1985}. A small value of the magnitude of $\Theta_m$ indicates that large changes in $p_m$ yield small changes in $u$. The estimation of parameter $p_m$ is therefore difficult in such case. When the sensitivity coefficient $\Theta_m$ is small, the inverse problem is ill-conditioned. If the sensitivity coefficients are linearly dependent, the inverse problem is also ill-conditioned. Therefore, to get an optimal evaluation of parameters $\vP$, it is desirable to have linearly-independent sensitivity functions $\Theta_m$ with large magnitude  for all parameters $p_m$. These conditions ensure the best conditioning of the computational algorithm to solve the inverse problem and thus the better accuracy of the estimated parameter.

It is possible to define the experimental design in order to reach these conditions. The issue is to find the optimal quantity of sensors $N^\circ$, their optimal location $\vX^\circ$, the optimal amplitude $A^\circ$ and the optimal frequency $\omega^\circ$ of the flux imposed at the surface $\Gamma_q$. To search this optimal experiment design, we introduce the following measurement plan:
\begin{align}\label{sec1_eq:measurement_plan}
  & \pi = \left\lbrace N, \vX, A, \omega \right\rbrace
\end{align}

In analysis of optimal experiment for estimating the unknown parameter(s) $\vP$, a quality index describing the accuracy of recovering is the $D-$optimum criterion \cite{Nenarokomov2005, Karalashvili2015, Artyukhin1985, Beck1977, Ucinski2004, VandeWouwer2000, Sun2005, Fadale1995, Emery1998, Anderson2005, Alifanov1995}:
\begin{align}\label{sec1_eq:D_optimum}
\Psi = \det \left[ F(\pi) \right]
\end{align}
where $F(\pi)$ is the normalized Fisher information matrix \cite{Karalashvili2015, Ucinski2004}:
\begin{subequations}\label{sec1_eq:fisher_matrix}
\begin{align}
& F(\pi) = \left[ \Phi_{ij} \right] &, \forall (i,j) \in \left\lbrace 1, \ldots, M \right\rbrace^2 \\
& \Phi_{ij} = \sum_{n=1}^N \int_0^{\tau} \Theta_i(x_n,t) \Theta_j(x_n,t) dt
\end{align}
\end{subequations}

The matrix $F(\pi)$ characterises the total sensitivity of the system as a function of measurement plan $\pi$ \eqs{sec1_eq:measurement_plan}. The search of the OED aims at finding a measurement plan $\pi^\star$ for which the objective function Eq. \eqref{sec1_eq:D_optimum} reaches the maximum value:
\begin{align}\label{sec1_eq:optimal_experimental_design}
& \pi^\circ = \left\lbrace N^\circ, \vX^\circ, A^\circ, \omega^\circ \right\rbrace = \arg \max_\pi \Psi 
\end{align}

To solve problem \eqref{sec1_eq:optimal_experimental_design}, a domain of variation $\Omega_{\,\pi}$ is considered for the quantity of sensors $N$, their location $\vX$, the amplitude $A$ and the frequency of the flux. Then, the following steps are done for each value of the measurement plan $\pi = \left\lbrace N, \vX, A, \omega \right\rbrace$ in domain $\Omega_{\,\pi}$.

The direct problem defined by\eqs{sec1_eq:direct_problem}(a-e) is computed. In this work, it is solved by using a finite-difference standard discretization. An embedded adaptive in time \textsc{Runge}--\textsc{Kutta} scheme combined with central spatial discretization is used. It is adaptive and embedded to estimate local error with little extra cost.

Given the solution $u$ of the direct problem \ref{sec1_eq:direct_problem}(a-e) for a fixed value of the measurement plan, the next step consists in computing $\Theta_m=\frac{\partial u}{\partial p_m}$ by solving the sensitivity problem associated to parameter $p_m$:
\begin{subequations}\label{sec1_eq:sensitivity_problem}
\begin{align}
& c \frac{\partial \Theta_m}{\partial t} - \frac{\partial }{\partial x} \left( k \frac{\partial}{\partial x} \Theta_m \right) =
- \frac{\partial c}{\partial p_m} \frac{\partial u}{\partial t} +
  \frac{\partial u}{\partial x} \frac{\partial}{\partial p_m} \left(\frac{\partial k}{\partial x} \right)  +
\frac{\partial k}{\partial p_m} \frac{\partial^2 u}{\partial xx}
&& x \in \Omega, t>0, \\
& \Theta_m=0  
&& x \in \Gamma_d , t>0, \\
& k \frac{\partial \Theta_m}{\partial x} = \frac{\partial u}{\partial x} \frac{\partial k}{\partial p_m}
&& x \in \Gamma_q , t>0, \\
& \Theta_m=0 
&& x \in \Omega,t=0, \\
& k=\left(k_0+k_1u \right)
\end{align}
\end{subequations}

The sensitivity problem \eqs{sec1_eq:sensitivity_problem} is also solved using an embedded adaptive in time \textsc{Runge}--\textsc{Kutta} scheme combined with central spatial discretization.

It is important to note that the solution of the direct \eqref{sec1_eq:direct_problem} problem and the sensitivity \eqref{sec1_eq:sensitivity_problem}(a-e) problem  are done for a given parameter $\vP$. The later is chosen with the prior information. The validity of the OED depends on the a priori knowledge. If there is no prior information, the methodology of the OED can be done using an outer loop on the parameter $\vP$ sampled using Latin hypercube or \textsc{Halton} quasi-random samplings for instance.

Then, given the sensitivity coefficients, the \textsc{Fisher} matrix \eqref{sec1_eq:fisher_matrix}(a,b) and the $D-$optimum criterion \eqref{sec1_eq:D_optimum} are calculated.


\subsection{Numerical example}
\label{sec1:numerical_example}

Current section illustrates the search of optimum experiment design for problem \eqs{sec1_eq:direct_problem} considering $u_0=u_D=1$. The dimension-less properties of the material are equal to $c^\star=10.9$, $k_0^\star=1$, $k_1^\star=0.12$. The final simulation time of the experiment is fixed to $\tau=28.3$.

\begin{figure}
\captionsetup{type=figure}
\begin{center}
\leavevmode
\subfloat[]{%
\label{sec1_fig:schema}
\includegraphics[width=0.48\textwidth]{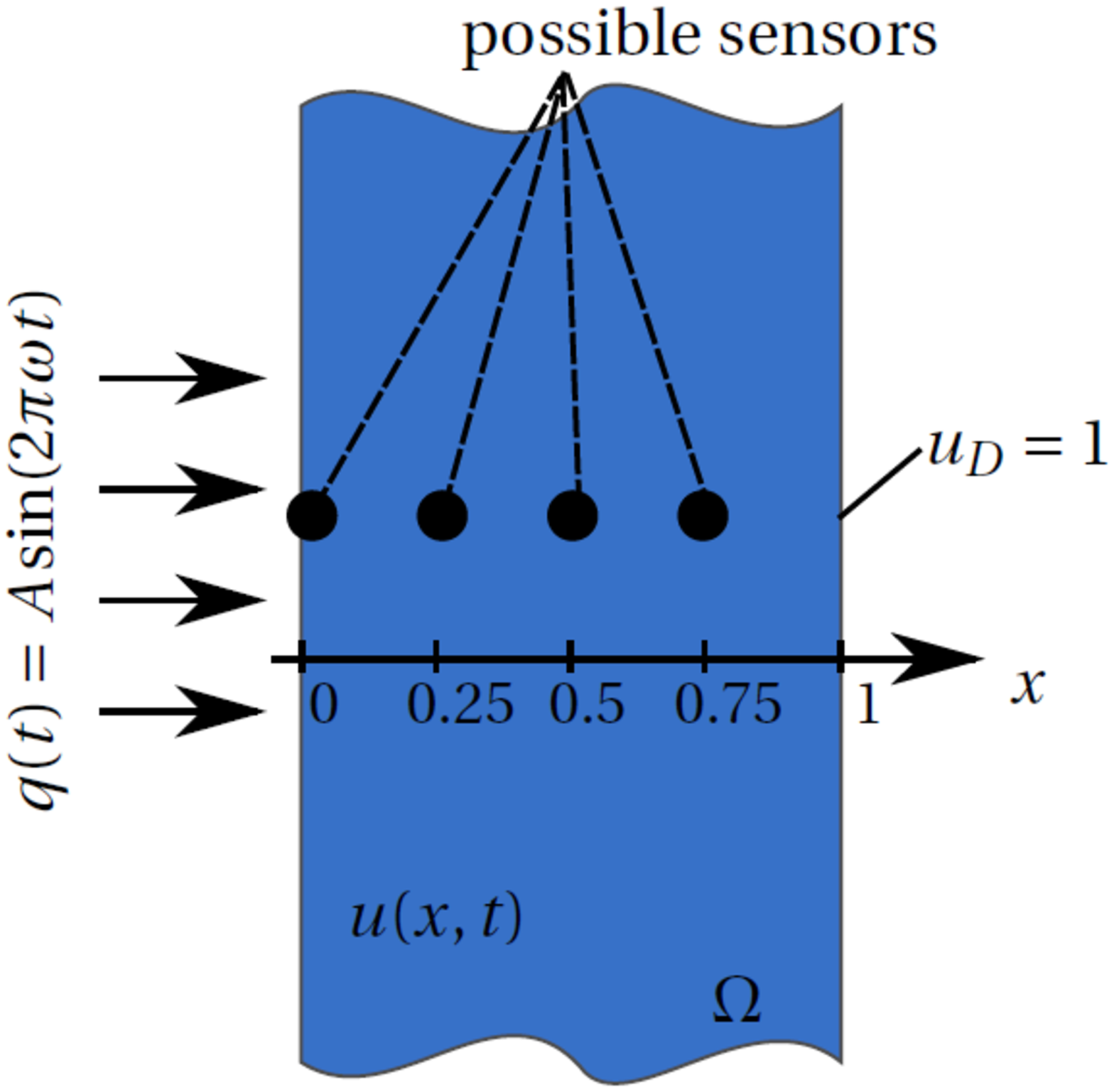}}
\hspace{0.4cm}
\subfloat[]{%
\label{sec1_fig:heat_flux}
\includegraphics[width=0.47\textwidth]{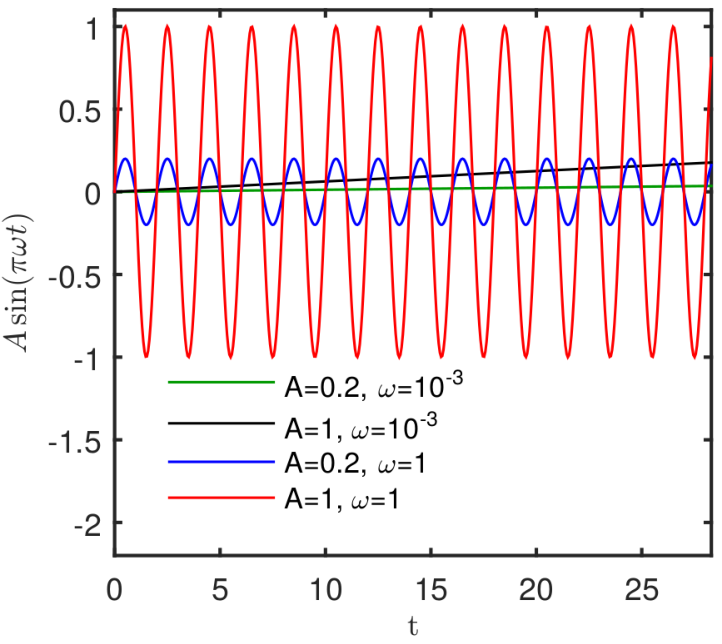}} 
\caption{\small\em Schematic view of experimental design (a) and the extreme values of the heat flux (b).}
\end{center}
\end{figure}

From a physical point of view, the numerical values correspond to a material of length $L_r=0.1$ m. The thermal properties were chosen from a wood fibre: $c=3.92 \cdot 10^5$ J/m$^3$/K, $k_0=0.118$ W/m/K  and $k_1=5 \cdot 10^{-4}$ W/m/K$^2$. The initial temperature of the material is considered uniform at $T_0=T_{ref}=293.15$ K. The temperature at the boundary $\Gamma_D$ is maintained at $T_D=T_{ref}=293.15$ K. The characteristic time of the problem is $t_{ref}=3050$ s. Thus, the total time of the experiments corresponds to 24 h. A schematic view of the problem is given in Figure~\ref{sec1_fig:schema}.

As mentioned in previous section, the OED is sought as a function of the quantity of sensors $N$, their location $\vX$, the amplitude $A$ and frequency $\omega$ of the flux imposed at the surface $\Gamma_q$. For the current numerical application, we consider a maximum of $N=4$ possible sensors. Their location varies from $\vX=\left[0 \right]$ for $N=1$ and $\vX=\left[0 \quad 0.25 \quad 0.5 \quad 0.75 \right]$ for $N=4$ as shown in Figure~\ref{sec1_fig:schema}. The variance of the error measurement equals $\sigma_T = 0.05^\circ$C. For the amplitude $A$, 5 values are considered in the interval $\left[0.2, 1\right]$ with a regular mesh of $0.2$. The minimal value of $A$ corresponds to a physical heat flux of 70 W/m$^2$. For the frequency, 30 values have been taken in the interval $\left[10^{-3}, 1\right]$. The extreme values of the heat flux are illustrated in Figure~\ref{sec1_fig:heat_flux}.

\subsubsection{Estimation of one single parameter}

\begin{figure}
\captionsetup{type=figure}
\begin{center}
\leavevmode
\subfloat[~IP($c$)]{%
\label{sec1_fig:surface_KSI_c}
\includegraphics[scale=1]{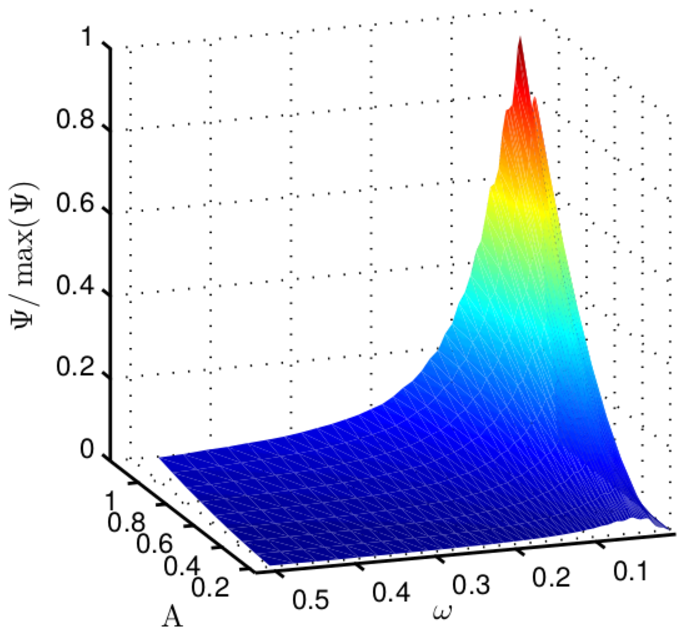}}
\subfloat[~IP($k_0$)]{%
\label{sec1_fig:surface_KSI_k0}
\includegraphics[scale=1]{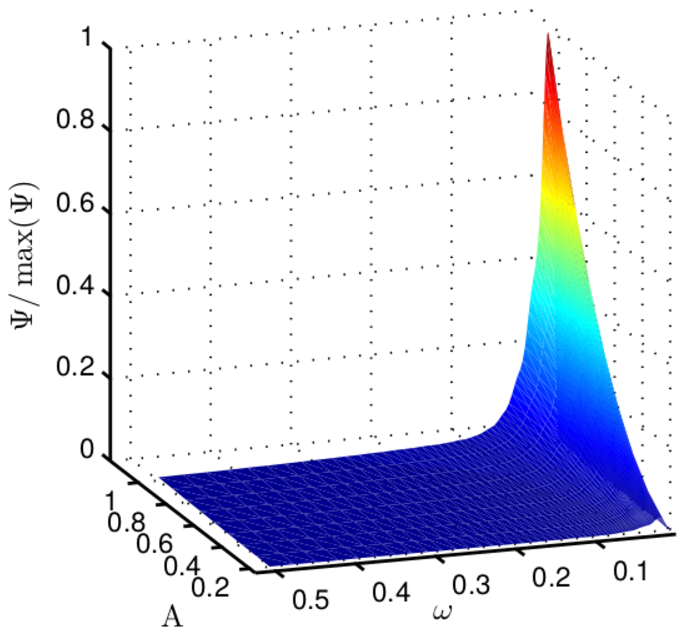}} \\
\subfloat[~IP($k_1$)]{%
\label{sec1_fig:surface_KSI_k1}
\includegraphics[scale=1]{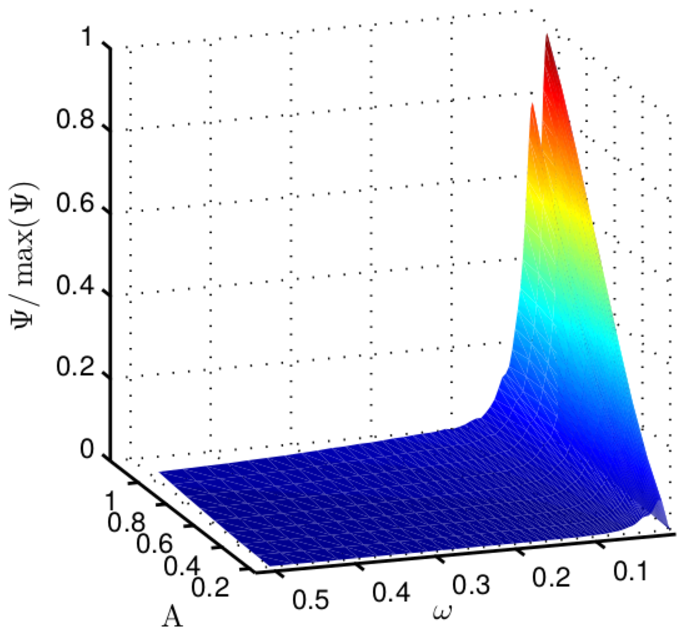}} 
\subfloat[~IP($(c,k_0,k_1)$)]{%
\label{sec1_fig:surf_ksi_multiparametre}
\includegraphics[scale=1]{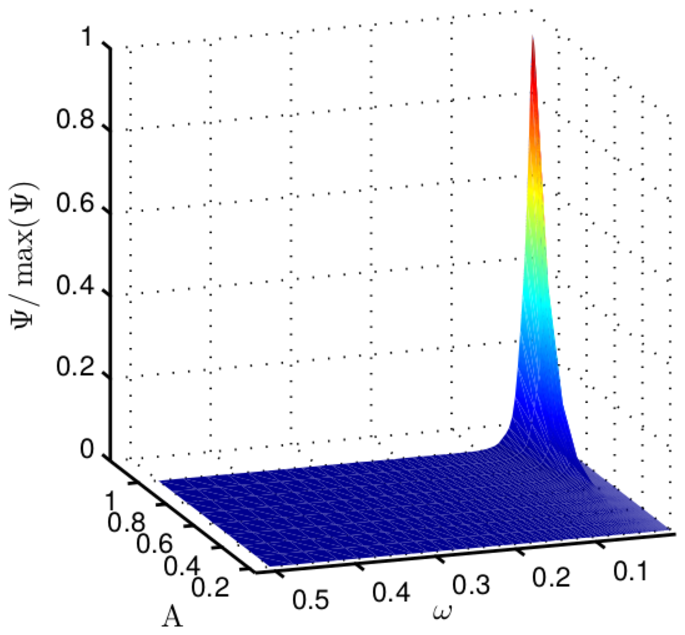}}
\caption{\small\em $D-$optimum criterion $\Psi$ as a function of $A$ and $\omega$ for the $4$ different experiments ($N\ =\ N^\circ\ =\ 1$).}
\end{center}
\end{figure}

\begin{figure}
\captionsetup{type=figure}
\begin{center}
\leavevmode
\includegraphics[width=0.69\textwidth]{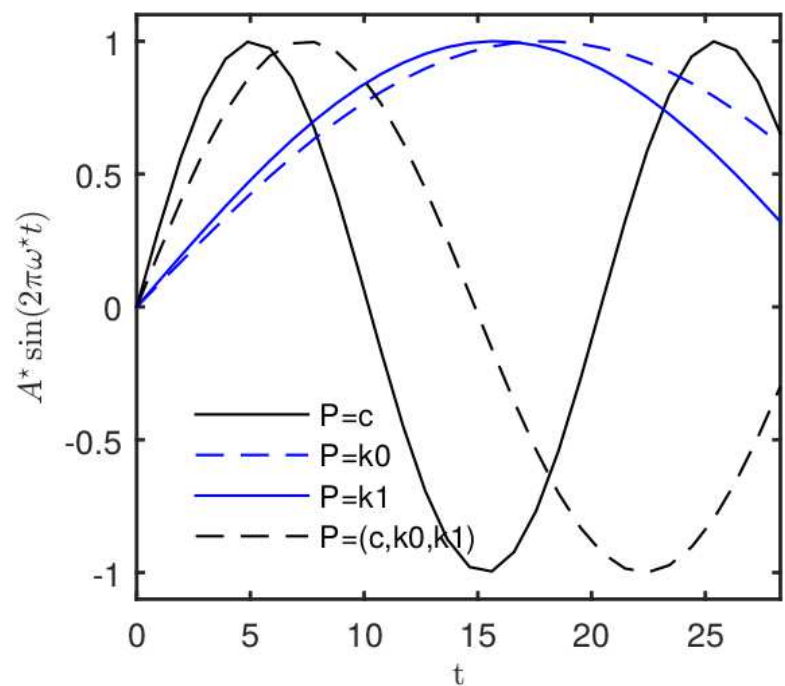}
\caption{\small\em Optimal heat flux for the four different experiments ($N\ =\ 1$).}
\label{sec1_fig:CL_star}
\end{center}
\end{figure}

Considering previous description of the numerical example, we first focus on the search of the OED for designing an experiment aiming at estimating one parameter. Thus, $\vP$ equals to $c$, $k_0$ or $k_1$. Figures~\ref{sec1_fig:surface_KSI_c}, \ref{sec1_fig:surface_KSI_k0} and \ref{sec1_fig:surface_KSI_k1} show the variation of the criterion $\Psi$ as a function of the amplitude $A$ and the frequency $\omega$ of the heat flux. For each of the three experiments, the criterion is maximum for a low frequency and for a large amplitude as presented in Figure~\ref{sec1_fig:CL_star}. Numerical values of the OED $\pi^\circ$ are given in Table~\ref{sec1_tab:resultats}. In terms of physical numerical values, the OED is reached for a heat flux of amplitude $350$ W/m$^2$ and a period of $17.3$ h, $60.6$h and $53.5$h for estimating parameter $c$, $k_0$ or $k_1$, respectively.

\begin{figure}
\captionsetup{type=figure}
\begin{center}
\leavevmode
\subfloat[IP($c$)]{%
\label{sec1_fig:KSI_c}
\includegraphics[width=0.48\textwidth]{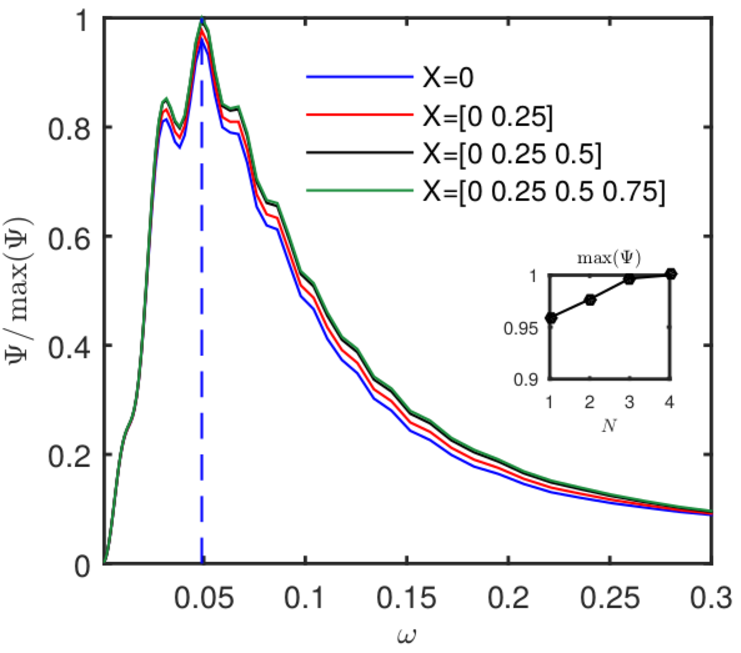}}
\subfloat[IP($k_0$)]{%
\label{sec1_fig:KSI_k0}
\includegraphics[width=0.48\textwidth]{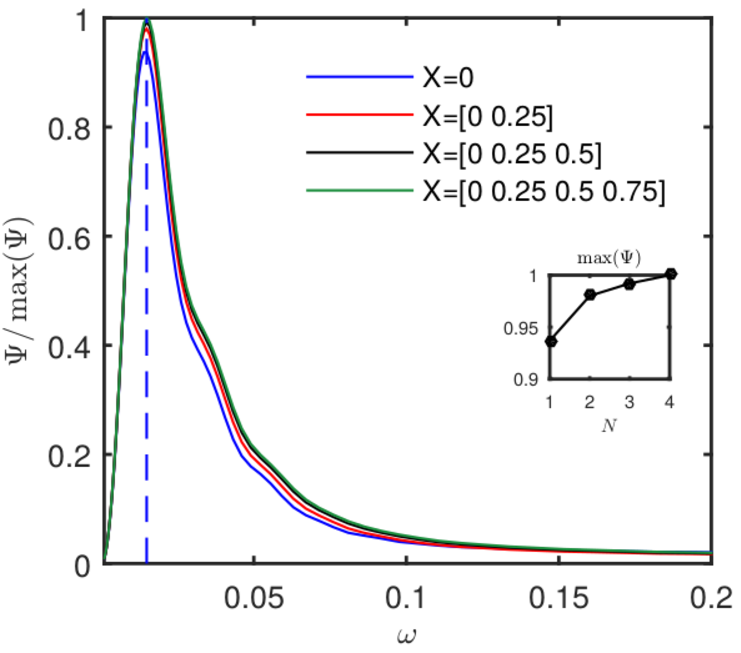}} \\
\subfloat[IP($k_1$)]{%
\label{sec1_fig:KSI_k1}
\includegraphics[width=0.48\textwidth]{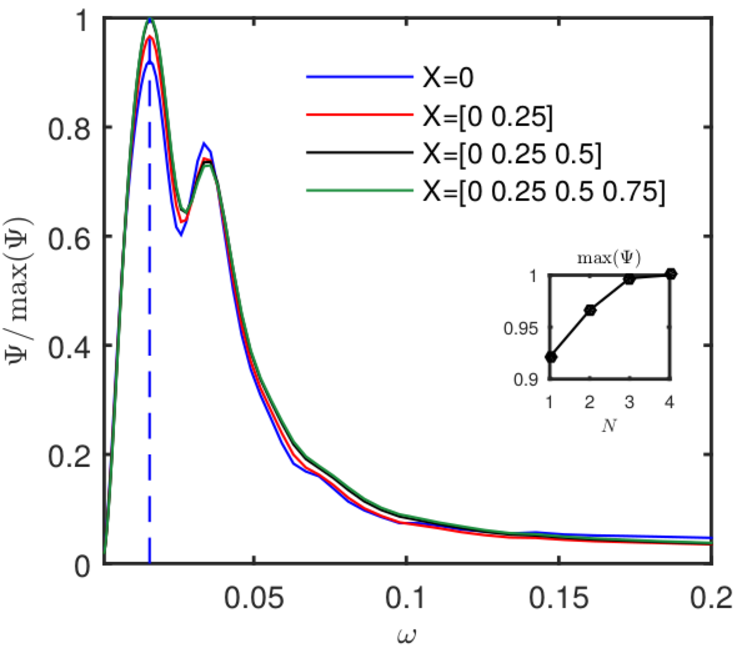}} 
\subfloat[IP$(c,k_0,k_1)$]{%
\label{sec1_fig:ksi_multiparametre}
\includegraphics[width=0.48\textwidth]{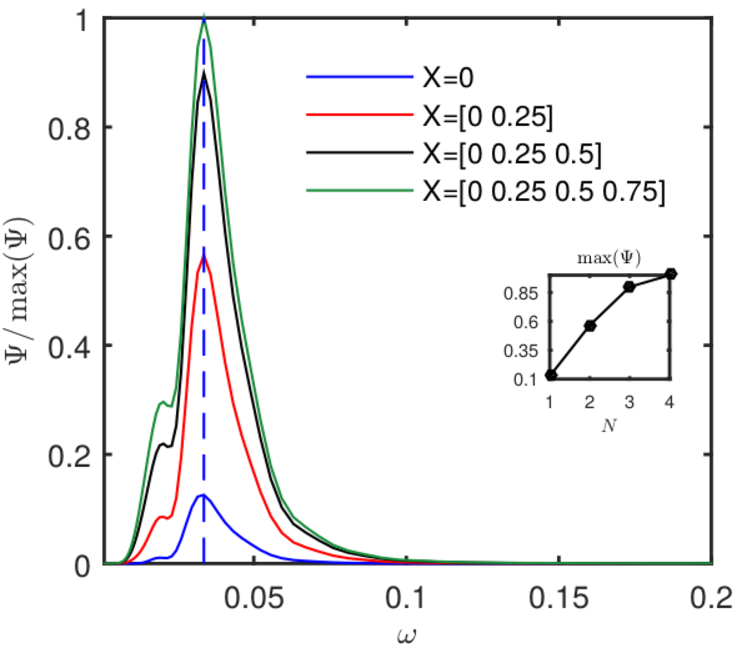}}
\caption{\small\em $D-$optimum criterion $\Psi$ as a function of $N$, $\vX$ and $\omega$ for the $4$ different experiments ($A\ =\ A^\circ\ =\ 1$).}
\end{center}
\end{figure}

\begin{figure}
\captionsetup{type=figure}
\begin{center}
\leavevmode
\subfloat[~IP$(c)$]{%
\label{sec1_fig:sens_c}
\includegraphics[width=0.48\textwidth]{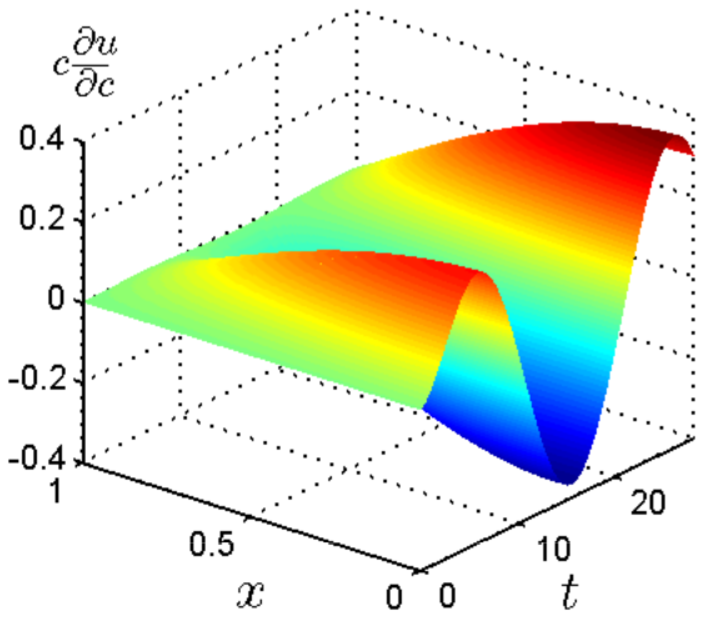}}
\subfloat[~IP$(k_0)$]{%
\label{sec1_fig:sens_k0}
\includegraphics[width=0.48\textwidth]{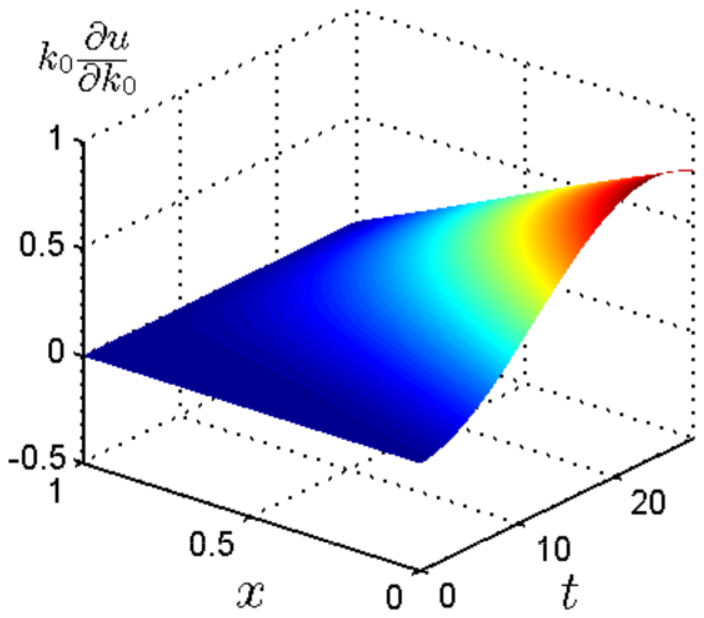}} \\
\subfloat[~IP$(k_1)$]{%
\label{sec1_fig:sens_k1}
\includegraphics[width=0.48\textwidth]{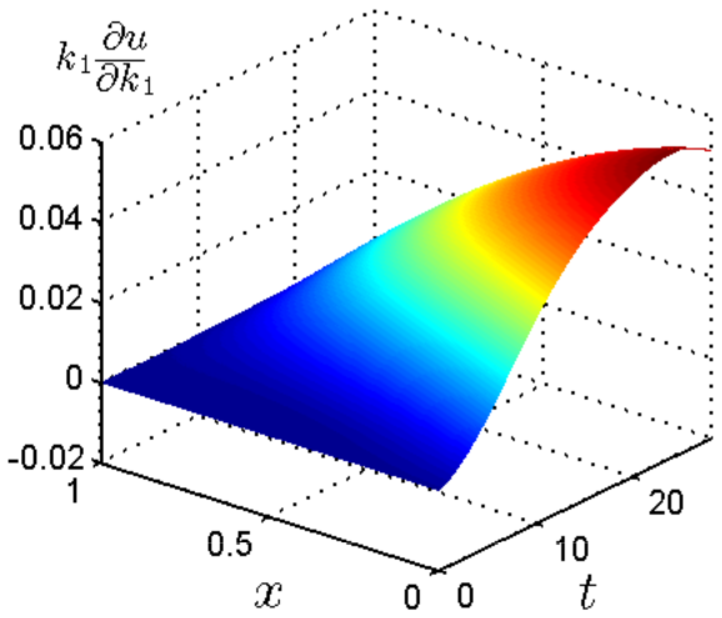}} 
\subfloat[~IP$(c,k_0,k_1)$]{%
\label{sec1_fig:ksi_multiparametre}
\includegraphics[width=0.48\textwidth]{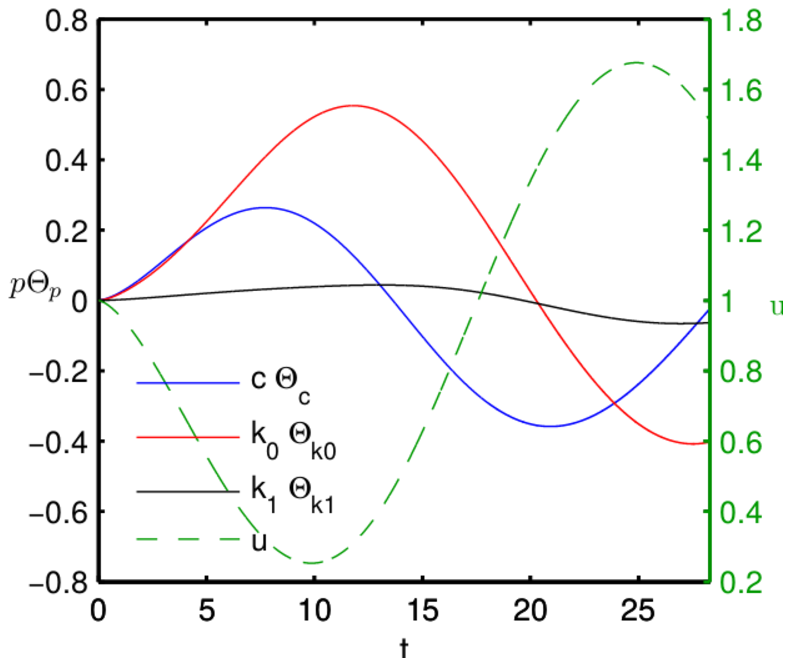}}
\caption{\small\em Sensitivity coefficient of parameters for the four different experiments ($N\ =\ N^\circ\ =\ 1$).}
\end{center}
\end{figure}

The effect of the numbers of sensors and their positions is given in Figures~\ref{sec1_fig:KSI_c}, \ref{sec1_fig:KSI_k0} and \ref{sec1_fig:KSI_k1} for a fixed amplitude. The criterion increases with the number $N$ of sensor. Adding new sensors yields to a more optimum design. However, it can be noticed the slope of the increase has a small magnitude. For a single sensor $N=1$, almost 95\% of the maximal criterion is reached. Therefore, if the amount of sensors for the experiments is limited, just one sensor located at the boundary receiving the heat flux is available reasonable, enabling to recover one parameter ($c$, $k_0$ or $k_1$). Indeed, the boundary $\Gamma_q$ is where the sensitivity coefficient of the parameter has the largest magnitude as shown in Figures~\ref{sec1_fig:sens_c}, \ref{sec1_fig:sens_k0} and \ref{sec1_fig:sens_k1} correspondingly.

It can be noticed in Figure \ref{sec1_fig:ksi_multiparametre} that the sensitivity coefficients on the surface $\Gamma_q \equiv \left\lbrace x=0 \right\rbrace$  are linearly-independent. Therefore, the inverse problem will not be very sensitive to measurement errors which enables an accurate estimation of the parameters $\vP=(c,k_0,k_1)$.

\begin{figure}
\captionsetup{type=figure}
\begin{center}
\leavevmode
\subfloat[]{%
\label{sec1_fig:sampling_k0k1}
\includegraphics[width=0.48\textwidth]{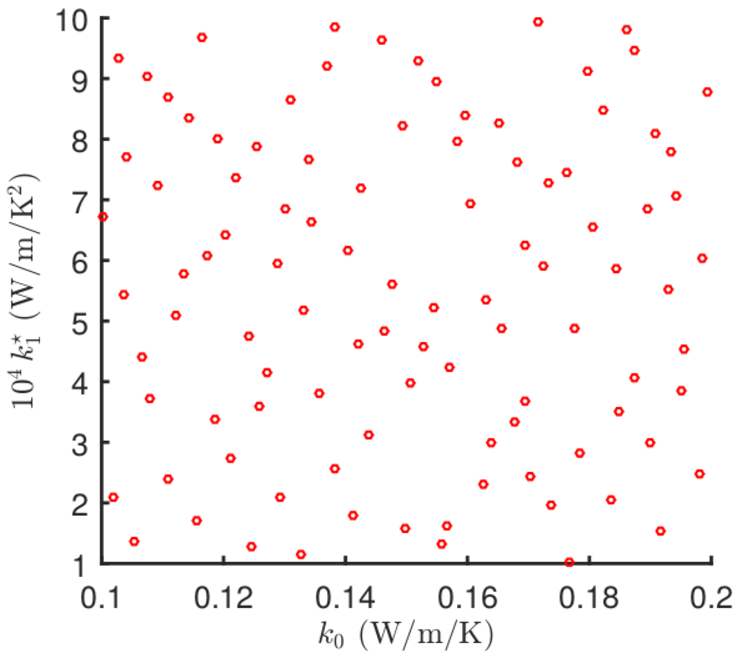}}
\subfloat[]{%
\label{sec1_fig:w_star_sampling}
\includegraphics[width=0.48\textwidth]{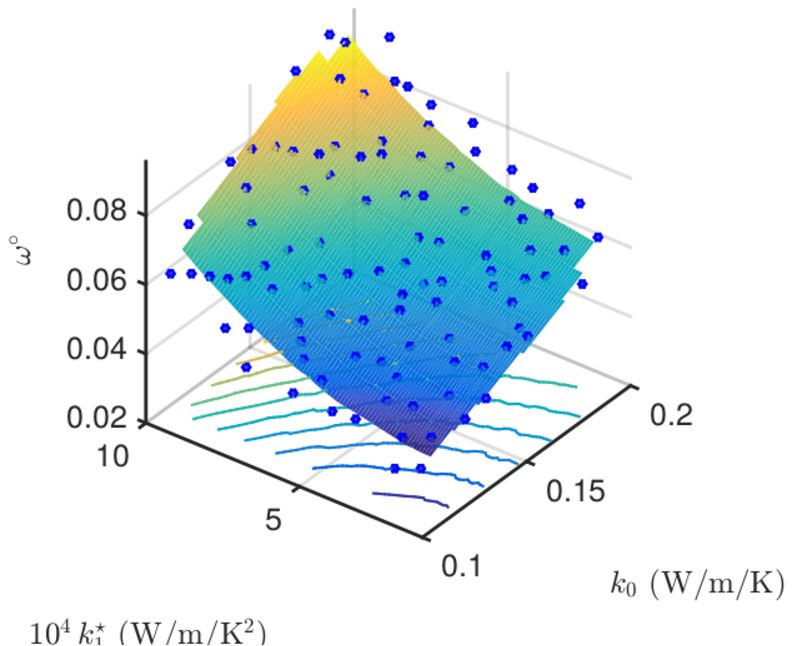}} 
\caption{\small\em Halton quasi-random sampling of parameters $(k_0,\,k_1)$ (a) and frequency optimal experiment design $\omega^\circ$ (b).}
\end{center}
\end{figure}

The OED is based on the solution of the direct problem~\ref{sec1_eq:direct_problem} and the sensitivity problem~\ref{sec1_eq:sensitivity_problem}, computed for a given value of the parameter $\vP$. The a-priori knowledge on parameter $\vP$ is crucial for the success of the OED. For some applications, there is no prior information on the parameter. For these cases, it is possible to operate an outer loop on the parameter $\vP$ sampled in a given domain. For this numerical example, it is considered that there is no prior information on parameters $k_0$ and $k_1$. It is only known their values belong to the domains  $\left[0.1, \, 0.2\right]$ W/m/K and  $\left[1 \, 10\right] \, 10^{-4}$ W/m/K$^2$. The a priori value of the volumetric heat capacity is still fixed to $c=3.92 \cdot 10^5$ J/m$^3$/K. A \textsc{Halton} quasi-random algorithm \cite{Kocis1997} is then used to generate a sampling of $100$ couples of parameters $(k_0,\,k_1)$ in the domains, as illustrated in Figure~\ref{sec1_fig:sampling_k0k1}. For each couple $(k_0,k_1)$, the methodology described in Section~\ref{sec1:OED} is performed to compute the OED. Figure~\ref{sec1_fig:w_star_sampling} gives the variation of optimal frequency $\omega^\circ$ with parameter $k_0$ and $k_1$. The blue points correspond to the results of the OED method for each couple of the discrete sampling. A surface of variation of $\omega^\circ$ is then interpolated. It can be noted that the increase of $\omega^\circ$ is more important with parameter $k_1$.

\begin{table}
\captionsetup{type=table}
\centering
\renewcommand*{\arraystretch}{1.25}
\begin{tabular}{@{}||>{\centering}m{.2\textwidth}||
>{\centering}m{.1\textwidth}||
>{\centering}m{.1\textwidth}
>{\centering}m{.1\textwidth}
>{\centering}m{.1\textwidth}
>{\centering}m{.1\textwidth}
>{\centering}m{.1\textwidth}
m{.1\textwidth}||}
\hline
\hline
\multirow{2}{*}{\small\hspace*{-0.75em} \textit{Unknown parameter}} & \multirow{2}{*}{$\max(\Psi)$} & 
\multicolumn{6}{c||}{\textit{Optimal experimental design} $\pi^\circ$} \\
  &   & $A^\circ$ (-) & $A^\circ$ (W/m$^2$) & $\frac{1}{\omega^\circ}$ (-)  & $\frac{1}{\omega^\circ}$  (h) & $N^\circ$ & $X^\circ$ \\
\hline
\hline
$\vP=c$ & $1.4 \e{-2}$ & 1 & 350 & 20.4 & 17.3 & 1 & 0 \\
$\vP=k_0$ & $8.6$ & 1 & 350 & 71.6 & 60.6 & 1 & 0 \\
$\vP=k_1$ & $3.23$ & 1 & 350 & 63.2 &  53.5 & 1 & 0 \\
$\vP=(c,k_0,k_1)$ & $9.1\e{-3}$ & 1 & 350 & 29.7 & 25.2 & 3 & 0 \\
\hline
\hline
\end{tabular}
\bigskip
\caption{\small\em Value of the maximum $D-$optimum criterion for the four different experiments.}
\label{sec1_tab:resultats}
\end{table}

\subsubsection{Verification of the ODE}

In order to illustrate the robustness of the ODE, several inverse problems are solved, considering different measurement designs. We consider $30$ inverse problems \eqref{sec1_eq:direct_problem} of parameter $\vP$ associated to the $30$ values of the frequency $\omega$ in the interval $\left[10^{-3}, 1\right]$. For a fixed value of $\omega$, $N_{\,e} \, = \, 100$ experiments were designed to recover the parameters $\vP$ equals to $c$, $k_0$ or $k_1$ and to $\vP=(c,k_0,k_1)$. For this purpose, the simulated discrete time observation is obtained by numerical resolution of the direct problem and a uniform sampling with time period $\Delta t \, = \, 10$ min. A normal distribution with a standard deviation of $\sigma \, = \, 0.01$ was assumed. The inverse problem is solved using Levenberg-Marquardt method \cite{Yang1999, Ozisik2000, Berger2016}. After the solution of the  $N_{\,e} \, \times \, 30 $ inverse problems, the empirical mean square error is computed:
\begin{align}\label{sec1_eq:EMSE}
\mathrm{EMSE}\left(\omega \right) \, = \, \sqrt{\frac{1}{N_{\,e}} \, \left(\vP \,-\, \vP^{\,\circ}\left(\omega \right) \right)^{\,2}}\,,
\end{align}
where $\vP^{\,\circ}$ is the estimated parameter by the resolution of the inverse problem.

The procedure is then repeated for the quantity $N$ and location $\vX$ of the sensors of the measurement plans and a fixed value of the frequency $\omega \, = \, \omega^{\,\circ}\,$. The empirical error is computed with equation \eqref{sec1_eq:EMSE} as a function of $N$ (as $\vP$ now depends on $N$). This approach is also applied for parameters $\vP$ equals to $c$, $k_0$ or $k_1$ and $\vP\ =\ (c,\,k_0,\,k_1)\,$.

\begin{figure}
\captionsetup{type=figure}
\begin{center}
\leavevmode
\subfloat[]{%
\label{sec1_fig:EMSE_N}
\includegraphics[width=0.48\textwidth]{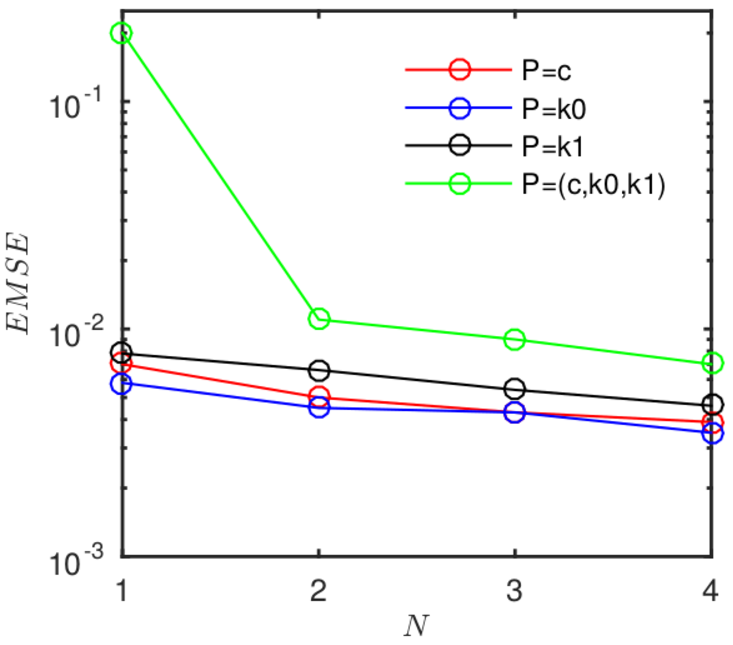}}
\subfloat[]{%
\label{sec1_fig:EMSE_w1}
\includegraphics[width=0.48\textwidth]{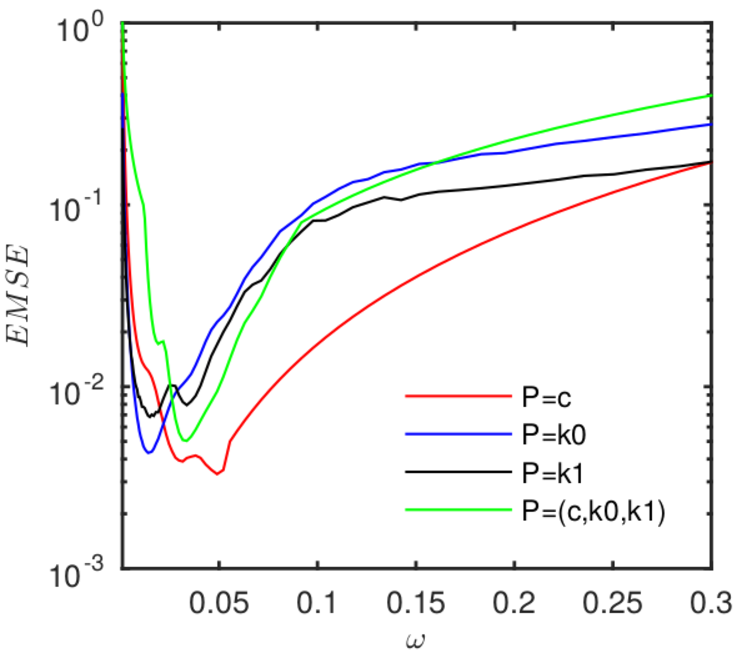}} 
\caption{\small\em Evolution of the empirical mean square error with the number of sensors (a) and with the frequency $\omega$ (b).}
\label{sec2_fig:EMSE}
\end{center}
\end{figure}

The evolution of the empirical mean square error is illustrated in Figures~\ref{sec1_fig:EMSE_N} and \ref{sec1_fig:EMSE_w1} for the parameter estimation problem corresponding to the four experiments.

In Figure~\ref{sec1_fig:EMSE_N}, the variation of the error with the number of the sensors is smooth for the experiments for the estimation of a single parameter. However, for the estimation of several parameters $\vP=(c,\,k_0,\,k_1)\,$, the EMSE is below $10^{-2}$ for three sensors. These results are in accordance with the variation of the criterion $\Psi$, in Figure~\ref{sec1_fig:ksi_multiparametre}. An important variation of the magnitude of $\Psi$ can be observed when passing from $\vX \,=\,0$ to $\vX \,=\,\left[0\, 0.25 \right]$.

In Figure~\ref{sec1_fig:EMSE_w1}, the error is minimised for the optimum experiment design $\omega^\circ$ for the four experiments. The estimation of the parameter tends to be inaccurate when the frequency tends to zero, as the heat flux tends to zero and there is almost no solicitation of the material.

For the experiments for the estimation of parameter $\vP=c$, the error variation is smooth. As illustrated in Figure~\ref{sec1_fig:EMSE_w1}, the peak of the criterion $\Psi$ has a large deviation. Therefore, if the frequency $\omega$ of the experiments is different from the one of the ODE $\omega^\circ$, the error to estimate the parameter might still be acceptable. For the other experiments, we can notice that the peak of the criterion $\Psi$ has a small deviation in Figures~\ref{sec1_fig:sens_k0}, \ref{sec1_fig:sens_k1} and \ref{sec1_fig:ksi_multiparametre}.

The peak of the criterion $\Psi$ has a small deviation in Figures~\ref{sec1_fig:sens_k0}, \ref{sec1_fig:sens_k1} and \ref{sec1_fig:ksi_multiparametre}. Consequently, the variation of the error in Figure~\ref{sec1_fig:EMSE_w1} is more important for estimating parameters $\vP\ =\ k_0$, $\vP\ =\ k_1$ or $\vP\ =\ (c,\,k_0,\,k_1)\,$. If the frequency of the experiment design is different from the one of the OED, it implies a loose of accuracy in the estimation of these parameters.

\section{Optimal Experiment Design for a non-linear coupled heat and moisture transfer problem}

In the previous section, the concept of optimal experiment design was detailed for a non-linear heat transfer problem. It has also been verified for $100$ inverse problems for different experiment designs by calculating the error as a function of the frequency and the number of sensors of the measurement plan. In next section, the approach goes further by studying optimal experiment design to estimate transport and storage coefficients of a heat and moisture transfer.

\subsection{Physical problem and mathematical formulation}

The physical problem concerns a 1-dimensional coupled heat and moisture transfer through a wall based on \cite{Rouchier2016, Berger2015, Tariku2010, Steeman2009}:
\begin{subequations}\label{sec2_eq:HAM}
\begin{align}
& c_{10} \frac{\partial T}{\partial t} 
- \frac{\partial }{\partial x} \left( d_{1} \frac{\partial T}{\partial x} \right) 
- \left(L_v -c_lT\right) \frac{\partial}{\partial x} \left( d_{2} \frac{\partial P_v}{\partial x} \right) = 0 \\
& \frac{\partial w}{\partial t} - \frac{\partial}{\partial x} \left( d_{2} \frac{\partial P_v}{\partial x} \right) = 0
\end{align}
\end{subequations}
with $w$ the water content, $P_v$ the vapour pressure, $T$ the temperature, $d_2$ the vapour permeability, $c_{10}$ the volumetric thermal capacity, $d_1$ the thermal conductivity, $c_l$ the specific heat capacity of water and $L_v$ the latent heat of evaporation. As this study remains in the hygroscopic range of the material, the liquid transport is not presently taken into account.

The following assumptions are adopted on the properties of the material. The volumetric moisture content is assumed as a first-degree polynomial of the vapour pressure. The vapour permeability and the thermal conductivity are taken as a first-degree polynomial of the volumetric moisture content:
\begin{subequations}
\label{sec2_eq:mat_hypothesis}
\begin{align}
& \frac{w}{w_0} = c_{20} +\frac{c_{21}}{w_0} P_v \\
& d_2 = d_{20} + d_{21} \frac{w}{w_0}\\
& k = d_{10} + d_{11} \frac{w}{w_0}
\end{align}
\end{subequations}

Based on these equations, the experimental set-up considers a material with uniform initial temperature and vapour pressure. At $t>0$, sinusoidal heat and vapour flux are imposed at boundary $\Gamma_q= \left\lbrace x=0 \right\rbrace$, while the temperature and vapour pressure are maintained constant at the other boundary $\Gamma_D= \left\lbrace x=1 \right\rbrace$. The unscaled problem can be formulated as:
\begin{subequations}
\label{sec2_eq:direct_problem}
\begin{align}
& c^\star_{10} \frac{\partial u}{\partial t^\star}
-\frac{\partial}{\partial x^\star} \left( d^{\star}_{1} 
\frac{\partial u}{\partial x^\star} \right)
- \left(\text{Ko}_1 - \text{Ko}_2 \right)\text{Lu} 
\frac{\partial}{\partial x^\star}  \left( d^{\star}_{2} 
\frac{\partial v}{\partial x^\star}\right) =0
&& x^{\star} \in \Omega, t^{\star} \in \left]0,\tau\right] \\[5mm]
& c^\star_{21} \frac{\partial v}{\partial t^\star} 
- \text{Lu} \frac{\partial}{\partial x^\star} \left(  d^{\star}_{2} 
\frac{\partial v}{\partial x^\star} \right) =0
&& x^{\star} \in \Omega, t^{\star} \in \left]0,\tau\right] \\[5mm]
& -d^{\star}_{1} \frac{\partial u}{\partial x^{\star}}  =A^{\star}_1 \sin(2\pi \omega^{\star}_1 t^{\star}) 
&& x^{\star}\in \Gamma_q , t^{\star}\in \left]0,\tau\right] \\[5mm]
& -d^{\star}_{2}\frac{\partial v}{\partial x^{\star}}  =A^{\star}_2 \sin(2\pi \omega^{\star}_2 t^{\star}) 
&& x^{\star} \in \Gamma_q , t^{\star}\in \left]0,\tau\right] \\[5mm]
& u=u_{D} , \quad v=v_{D} 
&& x^{\star} \in \Gamma_D , t^{\star}\in \left]0,\tau\right] \\
& u=u_0(x) , \quad v=v_0(x)
&& x^{\star} \in \Omega,t^{\star}=0 \\
& d_1^{\star} = d_{10}^{\star}+d_{11}^{\star}\left(c_{20} +c_{21}^{\star}v \right) \\
& d_2^{\star} = d_{20}^{\star}+d_{21}^{\star}\left(c_{20} +c_{21}^{\star}v \right)
\end{align}
\end{subequations}
with the following dimension-less ratios:
\begin{align*}
&  u  =\frac{T}{T_{ref}}
&& v  =\frac{P_v}{P_{ref}}
&& u_0=\frac{T_0}{T_{ref}}
&& v_0=\frac{P_{v,0}}{P_{ref}} \\[5mm]
& u_D=\frac{T_D}{T_{ref}} 
&& v_D=\frac{P_{v,D}}{P_{ref}}
&& \text{Ko}_1 = \frac{L_vc_{2,ref}}{c_{1,ref}}\frac{P_{ref}}{T_{ref}}
&& \text{Ko}_2 = \frac{c_Lc_{2,ref}}{c_{1,ref}}P_{ref} \\[5mm]
& \text{Lu} = \frac{d_{2,ref}c_{1,ref}}{c_{2,ref}d_{1,ref}}
&& d^{\star}_{10} = \frac{d_{10}}{d_{1,ref}} 
&& d^{\star}_{11} = \frac{d_{11}}{d_{1,ref}} 
&& d^{\star}_{20} = \frac{d_{20}}{d_{2,ref}}  \\[5mm]
& d^{\star}_{21} = \frac{d_{21}}{d_{2,ref}}  
&& c^{\star}_{10} = \frac{c_{10}}{c_{1,ref}} 
&& c^{\star}_{21} = \frac{c_{21}}{c_{2,ref}} 
&& c_{2,ref} = \frac{w_0}{P_{ref}} \\[5mm]
& A^{\star}_1 = \frac{A_1L}{d_{1,ref}T_{ref}}
&& A^{\star}_2 = \frac{A_2L}{d_{2,ref}P_{ref}}
&& \omega^{\star}_1 = \omega_1t_{ref}
&& \omega^{\star}_2 = \omega_2t_{ref}\\[5mm]
& t^\star = \frac{t}{t_{ref}} 
&& x^\star =\frac{x}{L} 
&& t_{ref} = \frac{c_{1,ref}L^2}{d_{1,ref}}
\end{align*}
where $\text{Ko}$ is the \textsc{Kossovitch} number, $\text{Lu}$ stands for the \textsc{Luikov} number, $L$ is the dimension of the material, $t_{ref}$, the characteristic time of the problem, $A$ and $\omega$ the amplitude and intensity of the heat and vapour fluxes. Subscripts $ref$ accounts for a reference value, $D$ for the \textsc{Dirichlet} boundary conditions, $0$ for the initial condition of the problem, $1$ for the heat transfer, $2$ for the vapour transfer and  superscript $\star$ for dimensionless parameters.


\subsection{Optimal experiment design}

The OED is sought as a function of the quantity of sensors $N$ and their locations $X$ and as a function of the frequencies $(\omega_1,\,\omega_2)$ of the heat and vapour fluxes. According to the results of Section~\ref{sec1:numerical_example} and to our numerical investigations, a monotonous increase of the sensitivity of the system were observed with the amplitude $(A_1,\,A_2)$ of the flux. Therefore, these parameters were considered as fixed. Thus, the OED aims at finding the measurement plan $\pi^\circ$ for which the criterion \eqs{sec1_eq:D_optimum} reaches a maximum value:
\begin{align}\label{sec2_eq:optimal_experimental_design}
& \pi^\circ = \left\lbrace N^\circ, \vX^\circ, \omega_1^\circ, \omega_2^\circ \right\rbrace = \arg \max_\pi \Psi 
\end{align}

Parameters  $L_v$, $c_L$ are physical constants given for the problem. Therefore, considering \eqs{sec2_eq:direct_problem}, a number of $7$ parameters can be estimated by the resolution of inverse problems: $(c^\star_{10},d^\star_{10}, d^\star_{11}, c^\star_{20},$ $c^\star_{21}, d^\star_{20}, d^\star_{21})$. One can focus on the definition of an experiment for the estimation of one single parameter or several parameters. It might be noted that parameters $(c^\star_{20}, c^\star_{21}, d^\star_{20}, d^\star_{21})$ can be identified by inverse problems considering field $u$, $v$ or both $(u,v)$ as observation. The thermal properties $(c^\star_{10},d^\star_{10}, d^\star_{11})$ can only be estimated using the observation of $u$.

All in all, $20$ experiments can be defined as:
\begin{enumerate}
\item[i.] 15 for the estimation of single parameters among $c^\star_{10}$, $d^\star_{10}$, $ d^\star_{11}$, $ c^\star_{20}$, $ c^\star_{21}$, $ d^\star_{20}$ or $ d^\star_{21}$,
\item[ii.] 1 for the estimation of the thermal properties $(c^\star_{10},d^\star_{10}, d^\star_{11})$,
\item[iii.] 3 for the estimation of the moisture properties $(c^\star_{20}, c^\star_{21}, d^\star_{20}, d^\star_{21})$, 
\item[iv.] 1 for the estimation of the hygrothermal properties (hg) $(c^\star_{10},d^\star_{10}, d^\star_{11}, c^\star_{20}, c^\star_{21}, d^\star_{20}, d^\star_{21})$.
\end{enumerate}
Following notation is adopted: $\text{IP}(p)\left[u\right] $ states for an experiment defined for the estimation of parameter $p$ using field $u$ as the observation. The 20 experiments are recalled in Table~\ref{sec2_tab:resultats}. The same methodology as presented in Section~\ref{sec1:OED_NL_heat} is used. The fifteen sensitivity functions are computed for calculating the criterion \eqref{sec2_eq:optimal_experimental_design}.


\subsection{Numerical example}
\label{sec2:NUM_example}

The following numerical values are considered for numerical application. The domain $\Omega$ is defined as $\Omega=\left[0,1\right]$, considering the wall thickness of the material as the characteristic length of the problem $L_r=0.1$m. The total simulation time of the experiments is $\tau=6 \e{3}$, corresponding to a physical simulation of $40$ days. The initial and prescribed conditions equal to $u_D=u_0=1$ and $v_D=v_0=0.5$. The reference temperature and vapour pressure are taken as $T_{ref}=293.15$K and $P_{v,ref}=2337$Pa, respectively. The amplitude of the heat and vapour fluxes are $A^\star_1=1.7\e{-2}$ and $A^\star_2=1.7$, equivalent to $600$W/m$^2$ and $1.2 \e{-7}$kg/m$^3$/s.

The dimension-less parameters are: 
\begin{align*}
& \text{Lu} = 2.5 \e{-4}
&& \text{Ko}_1 = 2.1 \e{-1}
&& \text{Ko}_2 = 2.5 \e{-2}
&& d^{\star}_{10} = 5 \e{-2}
&& d^{\star}_{11} = 5 \e{-3} \\
& d^{\star}_{20}= 1
&& d^{\star}_{21} = 0.4
&& c^{\star}_{10} = 1
&& c^{\star}_{20} = 2
&& c^{\star}_{21} = 6
\end{align*}
The properties correspond to a wood fibre material \cite{Rouchier2016, Berger2015}. They are given in its physical dimension in Appendix~A.

The OED is sought as a function  of the number of the sensors $N$. It varies from $N=1$, located at $X=\left[0\right]$, to $N=3$, located at $X=\left[0 \quad 0.2 \quad 0.4 \right]$. The variances of the measurement error are $\sigma_T=0.05 ^\circ$ and $\sigma_P=2$Pa. The OED is also investigated as a function of the frequencies $(\omega^\star_1,\omega^\star_2)$ of the flux. For each frequency, $20$ values are taken in the interval $\left[ 1 \e{-5};1.5 \e{-3} \right]$. The minimal and maximal values correspond to a flux having a physical period of $495$ days and $3.3$ days, respectively.

As mentioned in Section~\ref{sec1:OED}, the computation of the solution of the optimal experiment plan is done by successive iterations for the whole grid of the measurement plan $\pi=\left\lbrace N,X, \omega_1,\omega_2 \right\rbrace\,$.


\subsubsection{Estimation of one single parameter}

\begin{figure}
\captionsetup{type=figure}
\begin{center}
\leavevmode
\subfloat[]{%
\label{sec2_fig:KSIm_w1w2}
\includegraphics[width=0.48\textwidth]{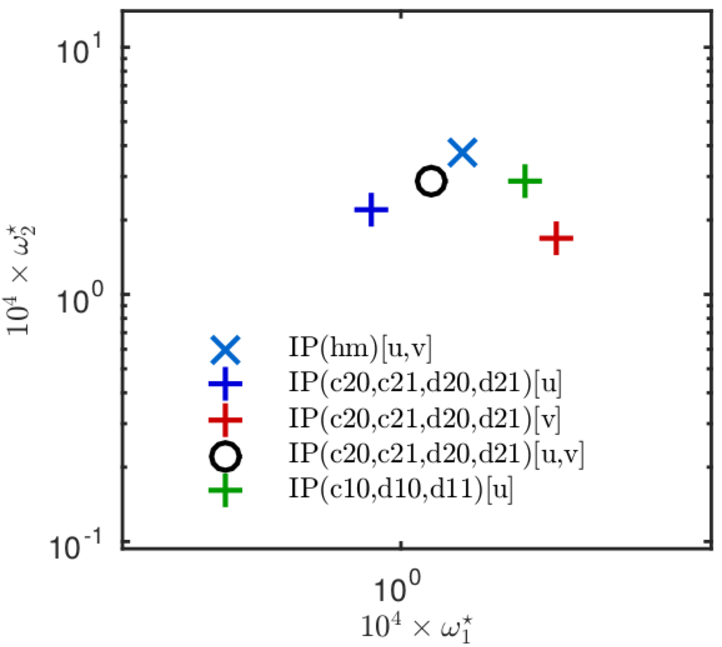}}
\subfloat[]{%
\label{sec2_fig:KSI_w1w2}
\includegraphics[width=0.48\textwidth]{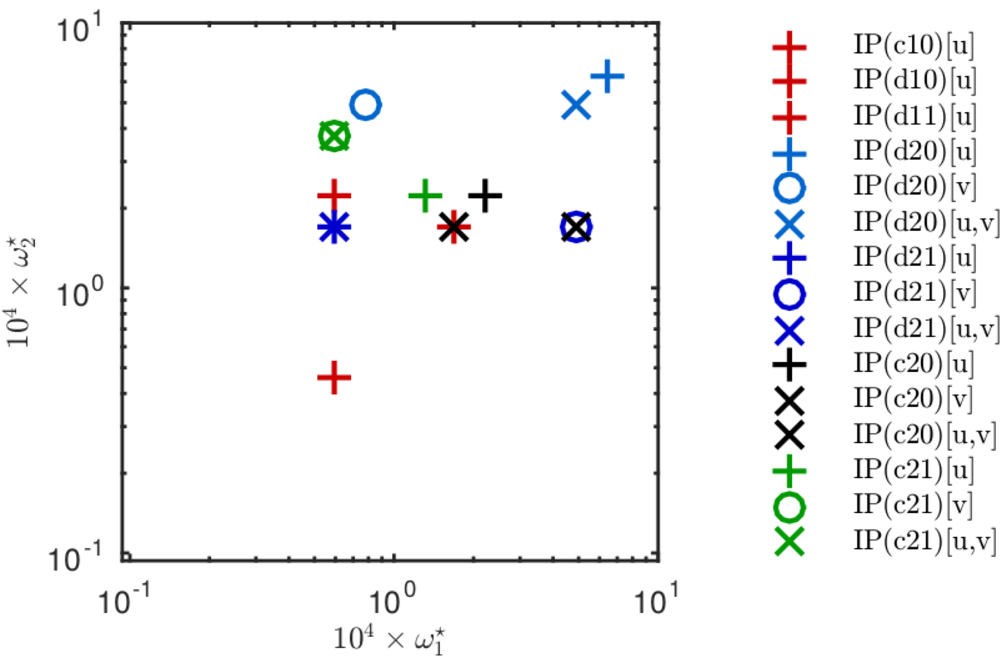}} 
\caption{\small\em Frequencies $(\omega^{\star \circ}_1,\,\omega^{\star \circ}_2)$ of the ODE.}
\label{sec2_fig:KSI_w1ow2o}
\end{center}
\end{figure}

In the current section, the OED is sought for experiments to estimate one single parameter among $c^\star_{10}$, $d^\star_{10}$, $ d^\star_{11}$, $ c^\star_{20}$, $ c^\star_{21}$, $ d^\star_{20}$ or $ d^\star_{21}$. The results of the ODE are given in Table~\ref{sec2_tab:resultats} for the physical values and in Figure~\ref{sec2_fig:KSI_w1w2} for the dimensionless values.

\begin{figure}
\captionsetup{type=figure}
\begin{center}
\leavevmode
\subfloat[~$\IP(c_{10})${[u]}, $N=1$]{%
\label{sec2_fig:surface_KSI_c10}
\includegraphics[width=0.48\textwidth]{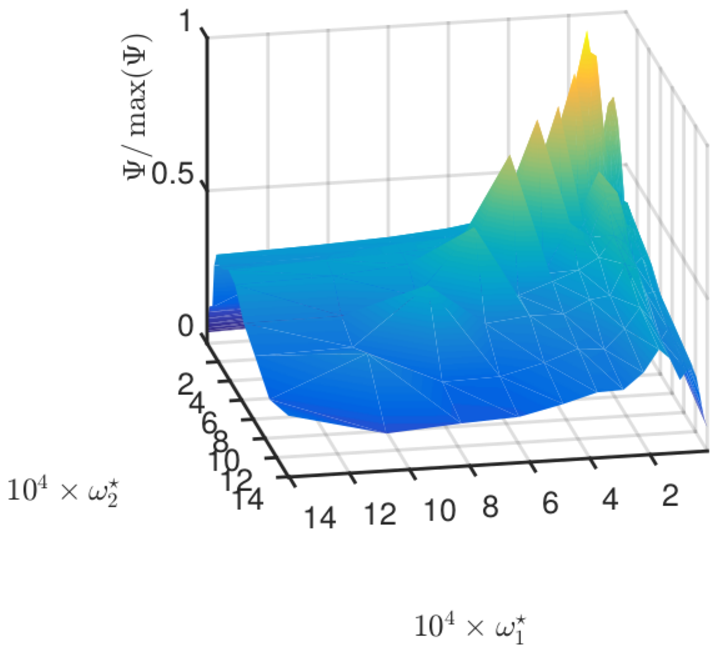}}
\subfloat[~$\IP(d_{10})${[u]}, $N=1$]{%
\label{sec2_fig:surface_KSI_d10}
\includegraphics[width=0.48\textwidth]{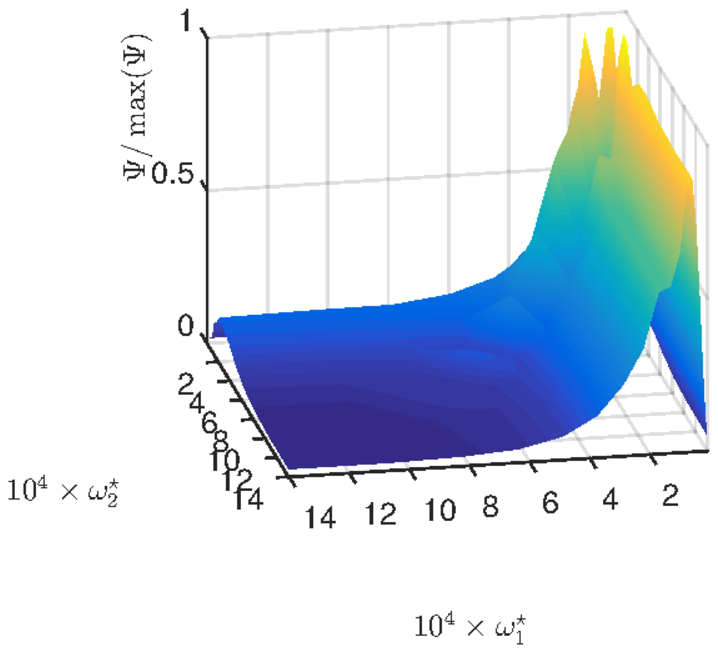}} \\
\subfloat[~$\IP(d_{11})${[u]}, $N=1$]{%
\label{sec2_fig:surface_KSI_d11}
\includegraphics[width=0.48\textwidth]{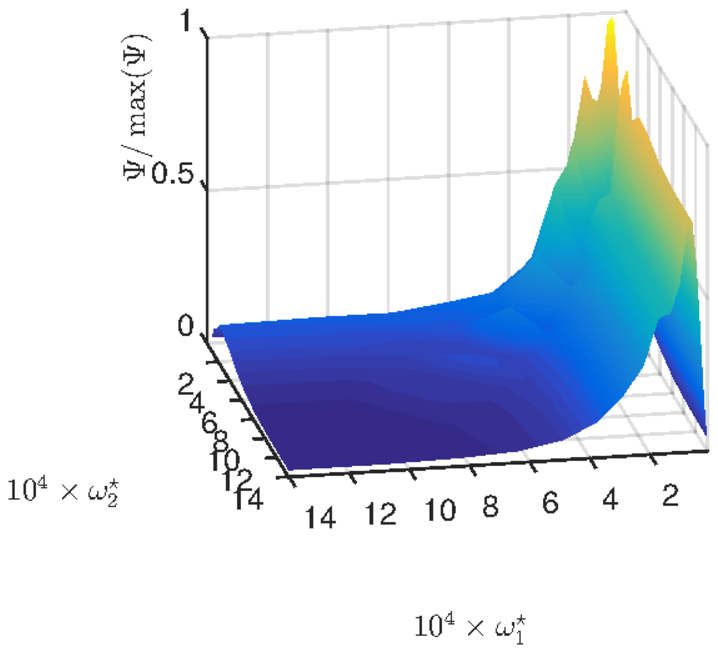}}
\subfloat[~$\IP(d_{20})${[u,v]}, $N=1$]{%
\label{sec2_fig:surface_KSI_d20}
\includegraphics[width=0.48\textwidth]{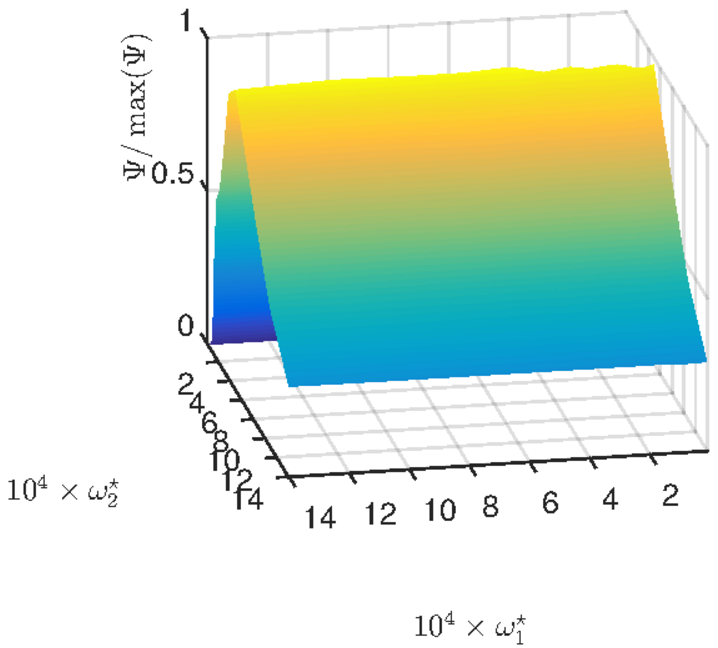}} \\
\subfloat[~$\IP(d_{21})${[u,v]}, $N=1$]{%
\label{sec2_fig:surface_KSI_d21}
\includegraphics[width=0.48\textwidth]{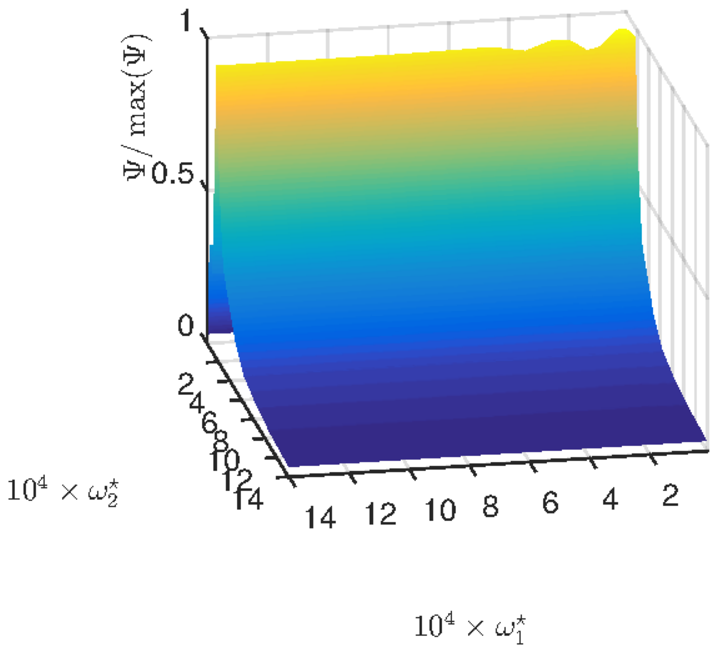}}
\subfloat[~$\IP(c_{20})${[u,v]}, $N=1$]{%
\label{sec2_fig:surface_KSI_c20}
\includegraphics[width=0.48\textwidth]{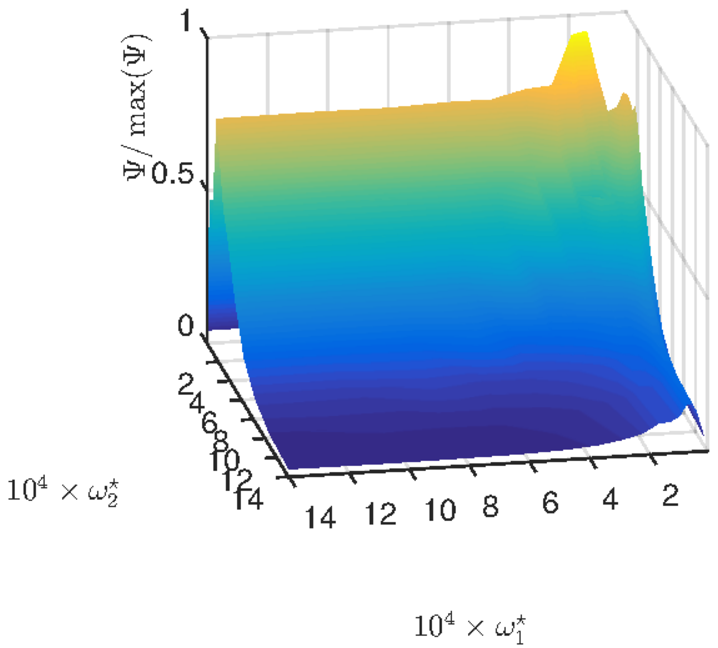}} \\
\caption{\small\em $D-$optimum criterion $\Psi$ as a function of the frequencies $(\omega_1,\omega_2)\,$.}
\label{sec2_fig:surface_KSI1}
\end{center}
\end{figure}

The criterion $\Psi$ varies actively with the frequencies $(\omega^\star_1,\,\omega^\star_2)$ for estimating parameter $c^\star_{10}\,$, as shown in Figure~\ref{sec2_fig:surface_KSI_c10}. The ODE is reached for a period for the heat and vapour flux of $27.2$ days.

On the other hand, Figures~\ref{sec2_fig:surface_KSI_d10} and \ref{sec2_fig:surface_KSI_d11} illustrate that the criterion varies mostly with the frequency $\omega^\star_1$ for estimating parameter $d^\star_{10}$ and $d^\star_{11}\,$. The ODE is reached for a period for the heat of $78.1$ days. Furthermore, the magnitude of $\Psi$ is really higher than zero, ensuring a good conditioning to solve inverse problems. As observed in the previous section concerning a non-linear heat transfer problem (Section~\ref{sec1:numerical_example}), the ODE period of the heat flux is shorter for the thermal capacity than for the thermal conductivity.

\begin{figure}
\captionsetup{type=figure}
\begin{center}
\leavevmode
\subfloat[]{%
\label{sec2_fig:surface_sensitivity_k0u_1}
\includegraphics[width=0.48\textwidth]{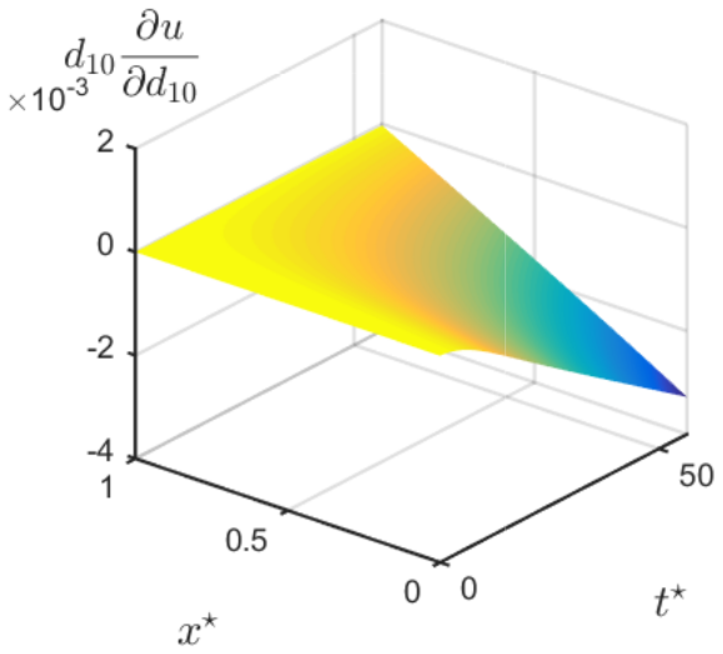}}
\subfloat[]{%
\label{sec2_fig:surface_sensitivity_k0u_2}
\includegraphics[width=0.48\textwidth]{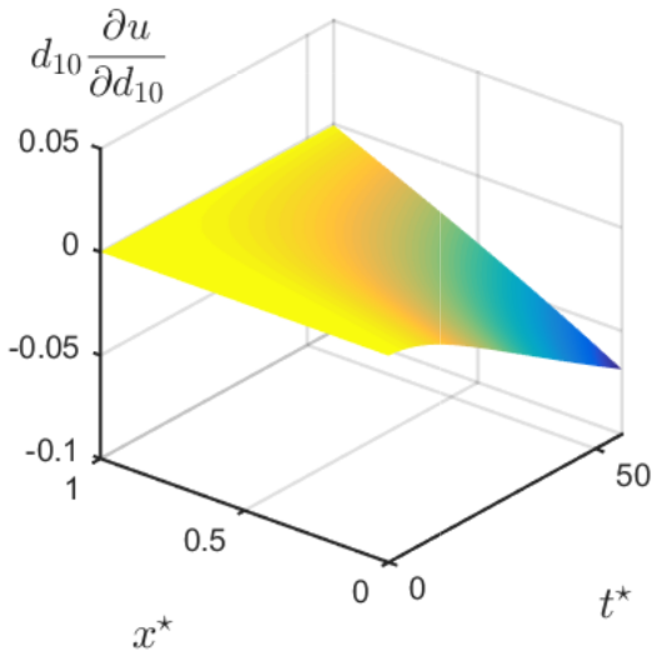}} 
\caption{\small\em Sensitivity coefficient of parameter $k_{0}$ for experimental conditions $\pi$ where $\Psi$ reaches its minimal value (a) ($\frac{2 \pi}{\omega_1}=3.3$ days, $\frac{2 \pi}{\omega_2}=495$ days, $N = 3$) and for the ODE conditions (b) ($\frac{2 \pi}{\omega_1^\circ}=78.1$ days, $\frac{2 \pi}{\omega_2^\circ}=20.9$ days, $N^\circ = 1$).}
\label{sec2_fig:surface_sensitivity_k0u}
\end{center}
\end{figure}

In Figure~\ref{sec2_fig:surface_sensitivity_k0u}, the sensitivity functions of parameters $d^\star_{10}$ is given for experimental conditions $\pi$ where $\Psi$ reaches its minimal value and for the ODE conditions (where $\Psi$ reaches it maximal value). In Figure~\ref{sec2_fig:surface_sensitivity_k0u_1}, the magnitude of the sensitivity function is almost 50 times smaller than one for the ODE conditions (Figure~\ref{sec2_fig:surface_sensitivity_k0u_2}). Therefore, the estimation of the parameters might be less accurate for this conditions than the other ones. In Figure~\ref{sec2_fig:surface_sensitivity_k0u_2}, it can be also noticed that the sensitivity function is maximal at the boundary $\Gamma_q\ =\ \left\lbrace x=0 \right\rbrace\,$. It emphasizes why the criterion $\Psi$ is maximal for a single sensor settled at this boundary.

For experiments estimating the vapour properties, $c^\star_{20}$, $c^\star_{21}$, $d^\star_{20}$ or $d^\star_{21}$, the ODE is not very sensitive to the frequency of the heat flux as reported in Figures~\ref{sec2_fig:surface_KSI_d20}, \ref{sec2_fig:surface_KSI_d21}, \ref{sec2_fig:surface_KSI_c20} and \ref{sec2_fig:surface_KSI_c21}. It can be noted that the criterion $\Psi$ is higher when dealing with experiments considering fields $(u,\,v)$ as observations. The computational algorithm to solve the inverse problem is better conditioned. The period of the ODE vapour flux is $9.5$ and $12.3$ days for experiments estimating $d^\star_{20}$ and $c^\star_{21}$. Experiments for parameters $d^\star_{21}$ and $c^\star_{20}$ have the same period of the ODE vapour flux ($27.2$ days).

\begin{figure}
\captionsetup{type=figure}
\begin{center}
\leavevmode
\subfloat[]{%
\label{sec2_fig:surface_sensitivity_c21u}
\includegraphics[width=0.48\textwidth]{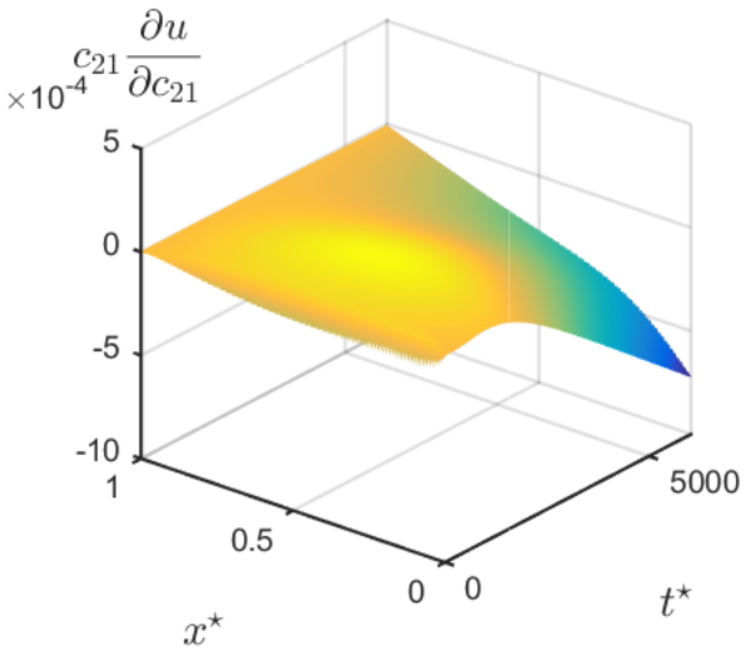}}
\subfloat[]{%
\label{sec2_fig:surface_sensitivity_c21v}
\includegraphics[width=0.48\textwidth]{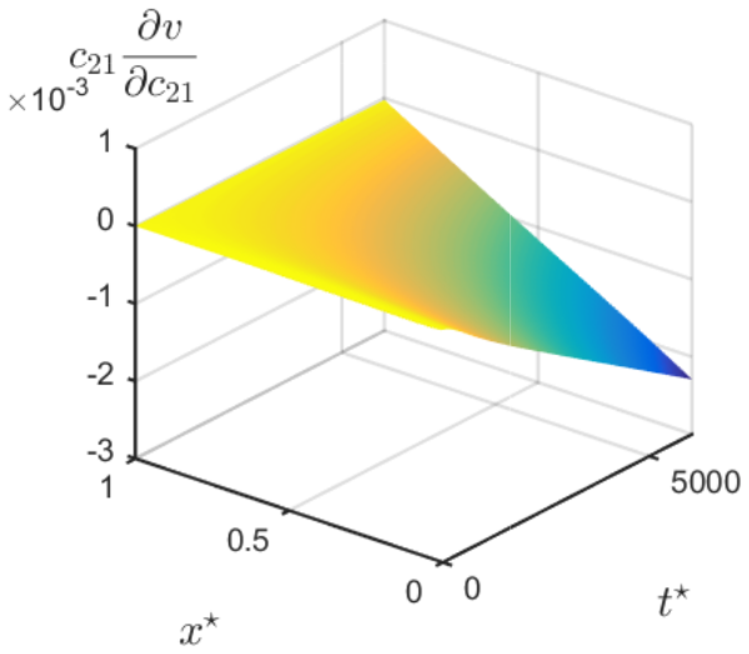}} 
\caption{\small\em Sensitivity coefficient of parameter $c_{21}$ for the ODE conditions of $\IP(c_{21})[u,v]\,$.}
\label{sec2_fig:surface_sensitivity_c21}
\end{center}
\end{figure}

It might be recalled that this analysis has been done for a fixed and constant error measurement, equals for the temperature and vapour pressure sensors. Indeed, this hypothesis can be revisited in practical applications. Furthermore, if only one field is available to estimate the vapour properties ($u$ or $v$), it is required to use the field $v$ as observation and prioritize the accuracy for those sensors. The criterion $\Phi$ and the sensitivity is highest for the field $v$ as shown in Figures~\ref{sec2_fig:surface_sensitivity_c21u} and \ref{sec2_fig:surface_sensitivity_c21v}.

For all experiments, a single sensor located at $x=0$ is sufficient to reach more than 95\% of the maximum criterion as given in Table~\ref{sec2_tab:resultats}. The surface receiving the heat and vapour flux is where the sensitivity of the parameters is the higher as illustrated in Figures~\ref{sec2_fig:surface_sensitivity_c21u} and \ref{sec2_fig:surface_sensitivity_c21v} for the parameter $c_{21}\,$.


\subsubsection{Estimation of several parameters}

The optimal experiment design is now sought for experiments to estimate several parameters: the thermophysical properties $(c^\star_{10},\, d^\star_{10},\, d^\star_{11})\,$, the moisture properties $(c^\star_{20},\, c^\star_{21},\, d^\star_{20},\, d^\star_{21})$ and the hygrothermal (hg) coefficients $(c^\star_{10},\, d^\star_{10},\, d^\star_{11},\, c^\star_{20},\, c^\star_{21},\, d^\star_{20},\, d^\star_{21})\,$. Five experiments are considered for the estimation of these parameters as reported in Table~\ref{sec2_tab:resultats}.

\begin{figure}
\captionsetup{type=figure}
\begin{center}
\leavevmode
\subfloat[~$\IP_{c_{21}}${[u,v]}, $N=1$]{%
\label{sec2_fig:surface_KSI_c21}
\includegraphics[width=0.48\textwidth]{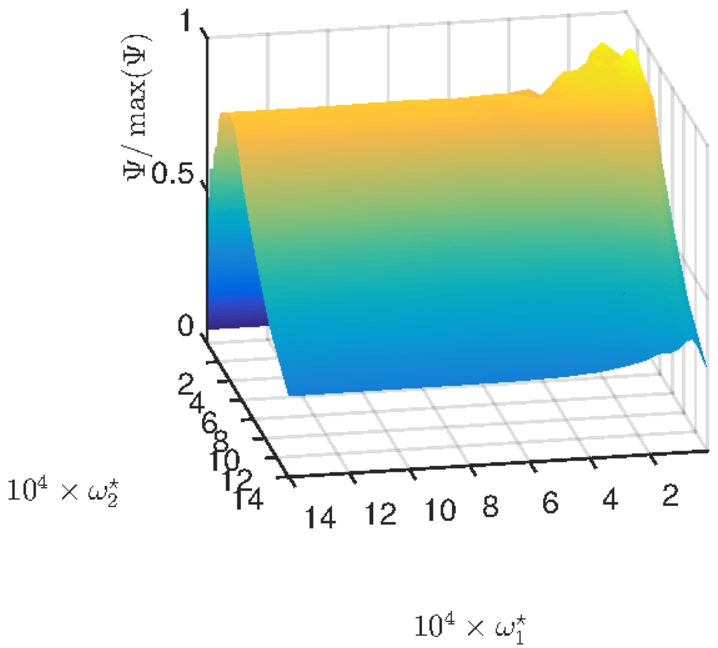}}
\subfloat[~$\IP(hg)${[u,v]}, $N=2$]{%
\label{sec2_fig:surface_KSI_hm}
\includegraphics[width=0.48\textwidth]{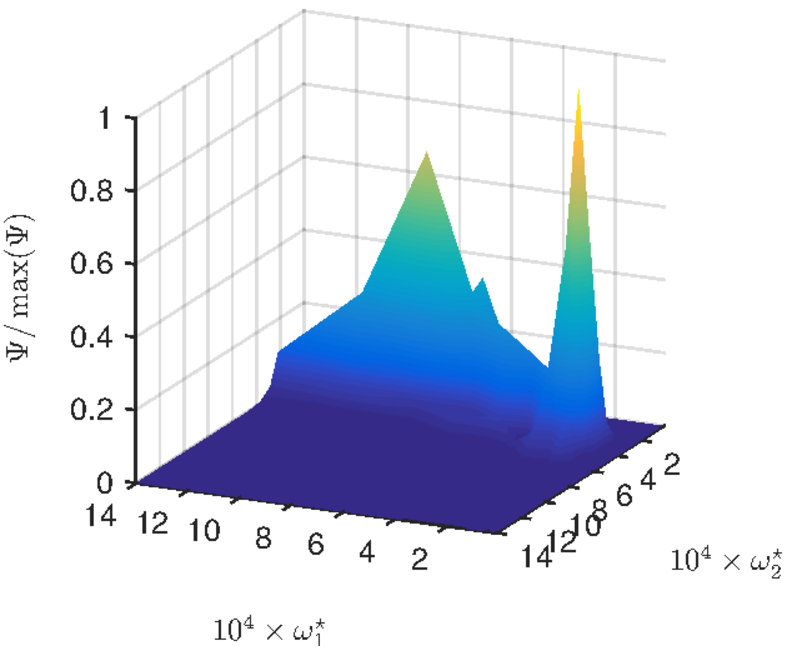}} \\
\subfloat[~$\IP(c_{20},c_{21},d_{20},d_{21})${[u,v]}, $N=1$]{%
\label{sec2_fig:surface_KSI_vapour}
\includegraphics[width=0.48\textwidth]{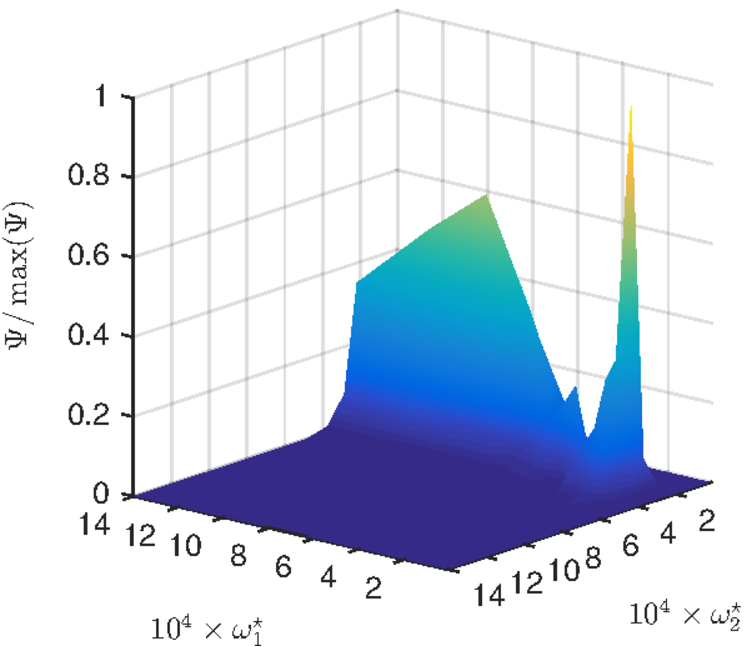}}
\subfloat[~$\IP(c_{10},d_{10},d_{11})${[u]}, $N=1$]{%
\label{sec2_fig:surface_KSI_heat}
\includegraphics[width=0.48\textwidth]{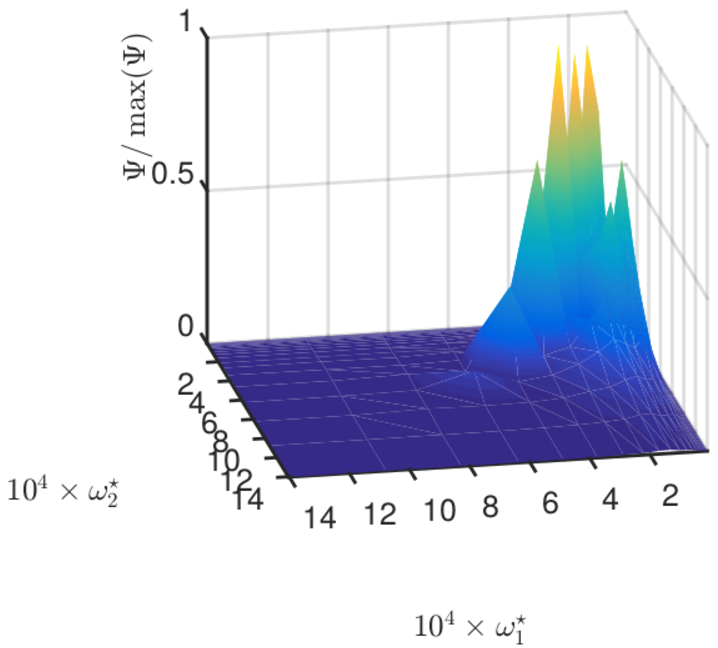}} \\
\caption{\small\em $D-$optimum criterion $\Psi$ as a function of the frequencies $(\omega_1,\,\omega_2)\,$.}
\label{sec2_fig:surface_KSI2}
\end{center}
\end{figure}

For the estimation of the hygrothermal properties, Figure~\ref{sec2_fig:surface_KSI_hm} shows that the criterion $\Psi$ varies mostly with frequency $\omega_1^\star\,$. The criterion is very close to zero (an order of $
\mathcal{O}(10^{-9})$). The computational algorithm for the solution of the inverse problem might be ill conditioned. The results of the inverse problem might not be accurate. A way to circumvent the problem is to increase the precision of the sensors and the amplitude of the heat and vapour fluxes.

According to Table~\ref{sec2_tab:resultats}, the moisture properties might be estimated using both fields $(u,\,v)\,$. If it is not possible, the field $v$ would give a more accurate estimation than field $u$. The criterion varies mostly with the frequency $\omega_1^\star$ of the heat flux, Figure~\ref{sec2_fig:surface_KSI_vapour}. The ODE is reached for a period of $35.4$ and $16$ days for the heat and vapour fluxes, respectively.

For the thermal properties, the criterion varies with both heat and vapour flux frequencies, as reported in Figure~\ref{sec2_fig:surface_KSI_heat}. A period of $16$ days for both fluxes yields the ODE.

\begin{figure}
\captionsetup{type=figure}
\begin{center}
\leavevmode
\includegraphics[width=0.99\textwidth]{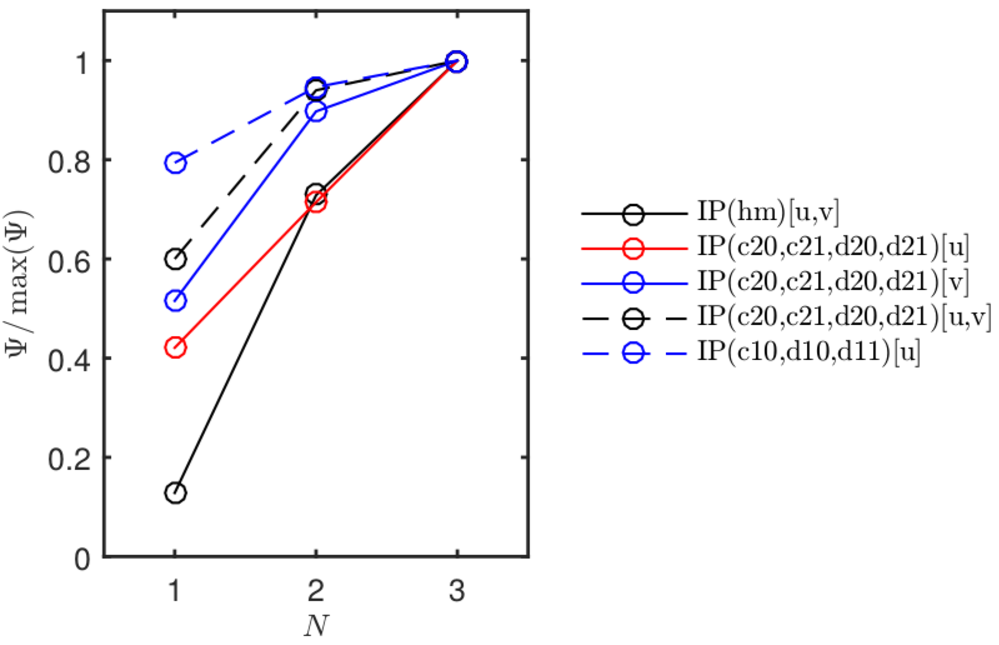}
\caption{\small\em $D-$optimum criterion $\Psi$ as a function of the quantity of sensors $N\,$.}
\label{sec2_fig:ksi_f_N}
\end{center}
\end{figure}

The variation of the criterion with the quantity of sensors is given in Figure~\ref{sec2_fig:ksi_f_N}. As expected, the criterion increases with the number of sensors for all experiments. For the estimation of the hygrothermal properties, the ODE is achieved for $3$ sensors. As the vapour properties might be estimated using both fields $(u,\,v)$, the use of two sensors, located at $x\ =\ 0$ and $x\ =\ 0.2$m, for each field, is a reasonable choice that enables to reach more than 95\% of the criterion $\Psi\,$. In the case, for any reasons, the measurement of both fields is not possible, three sensors measuring the field $v$ are required for the ODE. For the thermal properties, the use of only two sensors is reasonable, as 95\% of the maximum criterion is reached. These results are synthesised in Table~\ref{sec2_tab:resultats}.

\begin{table}
\captionsetup{type=table}
\centering
\renewcommand*{\arraystretch}{1.3}
\begin{tabular}{@{}||>{\centering}m{.3\textwidth}||
>{\centering}m{.1\textwidth}||
>{\centering}m{.1\textwidth}
>{\centering}m{.1\textwidth}
m{.1\textwidth}||}
\hline
\hline
\multirow{2}{*}{\textit{Unknown parameter}} & \multirow{2}{*}{$\max \left\lbrace\Psi \right\rbrace$} & 
\multicolumn{3}{c||}{\textit{Optimal experimental design} $\pi^\star$} \\
  &   & $ \displaystyle  \frac{2 \pi}{\omega_1^{\circ}}$ (days) & $ \displaystyle  \frac{2 \pi}{\omega_2^\circ}$ (days) & $N^\circ$ \\
\hline
\hline
$\IP(c_{10})[u]$  & 6.03 & 27.2 & 27.2  & 1 \\
$\IP(d_{10})[u] $ & 288    & 78.1 & 20.9  & 1 \\
$\IP(d_{11})[u]$  & 181    & 78.1 & 101.7 & 1 \\
$\IP(d_{20})[u]$  & 0.53   & 7.3  & 7.3   &  1 \\
$\IP(d_{20})[v]$  & 0.99   & 60   & 9.5   & 1 \\
$\IP(d_{20})[u,v]$& 1.53   & 9.5  & 9.5   & 1 \\
$\IP(d_{21})[u]$  & 133    & 78.1 & 27.2  & 1 \\
$\IP(d_{21})[v]$  & 140    & 9.5  & 27.2  & 1 \\
$\IP(d_{21})[u,v]$& 276    & 78.1 & 27.2  & 1 \\
$\IP(c_{20})[u]$  & 0.02   & 20.9 & 20.9  & 1 \\
$\IP(c_{20})[v]$  & 0.03   & 9.5  & 27.2  & 1 \\
$\IP(c_{20})[u,v]$& 0.05   & 27.2 & 27.2  & 1 \\
$\IP(c_{21})[u]$  & 0.004  & 35.4 & 20.9  & 1 \\
$\IP(c_{21})[v]$  & 0.014  & 78.1 & 12.3  & 1 \\
$\IP(c_{21})[u,v]$& 0.017  & 78.1 & 12.3  & 1 \\
\hline
$\IP(hg)[u,v]$                   & $4.5 \e{-9}$     & 27.2 & 12.3 & 3 \\
$\IP(c_{20},c_{21},d_{20},d_{21})[u]$  & 0.001    & 60.0 & 20.9 & 3 \\
$\IP(c_{20},c_{21},d_{20},d_{21})[v]$  & 0.15     & 12.3 & 27.2 & 3 \\
$\IP(c_{20},c_{21},d_{20},d_{21})[u,v]$& 0.2      & 35.4 & 16 & 2 \\
$\IP(c_{10},d_{10},d_{11})[u]$       & 137    & 16 & 16 & 2 \\
\hline
\hline
\end{tabular}
\bigskip
\caption{\small\em Value of the maximum $D-$optimum criterion for each experiments.}
\label{sec2_tab:resultats}
\end{table}

\section{Conclusions}

In the context of estimating material properties, using in-site measurements of wall in test cells or real buildings combined with identification methods, this study explored the concept of optimal experiment design (OED). It aimed at searching the best experimental conditions in terms of quantity and location of sensors and flux imposed to the material. These conditions ensure to provide the best accuracy of the identification method and thus the estimated parameter. The search of the OED was done using the \textsc{Fisher} information matrix, quantifying the amount of information contained in the observed field.

Two cases were illustrated. The first one dealt with an inverse problem of non-linear heat transfer to estimate the thermal properties (storage and transport coefficients), considering a uniform initial temperature field and, as boundary conditions, a fixed prescribed temperature on one side and a sinusoidal heat flux on the other one, for $24$ hours. The ODE yields in using one single temperature sensor located at the surface receiving the flux. The flux should have an intensity of $350$ W/m$^2$ and experiment periods of $17$ h, $61$ h, $54$ h and $25$ h for the estimation of thermal capacity, thermal conductivity, temperature dependent conductivity and all parameters, respectively. For this case study, the concept of optimal experiment was verified by solving $100$ inverse problems for different experiment designs and calculating the error as a function of the frequency and the number of sensors of the measurement plan. It has been demonstrated that the accuracy of the parameter is higher when parameters are recovered with measurements carried out via ODE.

The second application concerned experiments for inverse problems of a coupled non-linear heat and moisture transfer problem to estimate the hygrothermal properties of the material. The experiment is similar to the first case study. Uniform initial distribution of temperature and vapour pressure fields were considered, submitted to harmonic heat and vapour fluxes at one side and prescribed temperature and vapour pressure values at the other one. The experiment was done for a period of $40$ days. The achievement of the ODE was explored for different experiments, aiming at estimating one or several parameters. As the equation considered are weakly coupled, the thermal properties can only be determined using the temperature. The accuracy of the identification method does not depend on the vapour flux. For the vapour properties, results have shown that the estimation will be more accurate using the temperature and vapour pressure as observations. Furthermore, the accuracy actively depends on the period of the vapour flux. Single sensor has to be located at the side where the flux is imposed. For experiments to estimate all the hygrothermal properties, two sensors are enough to improve the accuracy.

This contribution explored the concept of optimal experiment design for application in building physics for estimating the hygrothermal properties of construction materials. The methodology of searching the ODE is important before starting any experiment aiming at solving parameter estimation problems. With a priori values of the unknown parameters, the sensitivity functions and the optimum criterion can be computed. Results allow choosing by means of deterministic approach the conditions of the experiments. A good design of experiments avoids installing unnecessary sensors. In the case of coupled phenomena, as the combined heat and moisture transfer problem, considering sensor accuracies, ODE enables to choose and select the field that must be monitored. It also improves the accuracy of the solution of the estimation problem.

Further work is expected to be carried out with different design strategies (ODE and others), estimating properties using real observations.


\bigskip

\subsection*{Acknowledgments}
\addcontentsline{toc}{subsection}{Acknowledgments}

The authors acknowledge the Brazilian Agencies \texttt{CAPES} of the Ministry of Education and the \texttt{CNPQ} of the Ministry of Science, Technology and Innovation for the financial support. Dr.~\textsc{Dutykh} also acknowledges the hospitality of \texttt{PUCPR} during his visit in April 2016.

\bigskip


\appendix
\section{Hygrothermal properties}
\label{annexe:HAM_prop}

The hygrothermal properties of the material used in Section~\ref{sec2:NUM_example} are given in Table~\ref{Anx_tab:HAM_prop}.

\begin{table}
\centering
\renewcommand*{\arraystretch}{1.3}
\begin{tabular}{||>{\centering}m{.3\textwidth}||>{\centering}m{.3\textwidth}||}
\hline
\hline
\textit{Properties} & \textit{Value}
\tabularnewline
\hline
$d_{10}$ (W/m/K) & $0.5$
\tabularnewline
$d_{11}$ (W/m/K/Pa) & $0.05$
\tabularnewline
$d_{20}$ (s) & $2.5 \e{-11}$
\tabularnewline
$d_{21}$ (s/Pa) & $1 \e{-11}$
\tabularnewline
$c_{11}$ (J/m$^3$/K) & $4 \e{5}$
\tabularnewline
$c_{20}$ (-) & $2$
\tabularnewline
$c_{21}$ (s$^2$/m$^2$) & $2.5 \e{-2}$
\tabularnewline
\hline
\textit{Physical constant} & \textit{Value}
\tabularnewline
\hline
$L_v$ (J/kg) & $2.5 \e{6}$
\tabularnewline
$c_{L}$ (J/kg) & $1000$
\tabularnewline
\hline
\hline
\end{tabular}
\bigskip
\captionsetup{type=table}
\caption{\small\em Hygrothermal properties of the material.}
\label{Anx_tab:HAM_prop}
\end{table}


\addcontentsline{toc}{section}{References}
\bibliographystyle{abbrv}
\bibliography{biblio}

\end{document}